\documentclass{llncs}

\usepackage{amssymb}
\usepackage{amsmath}
\usepackage{stmaryrd}
\usepackage{mathrsfs}
\usepackage{makeidx}
\usepackage{graphicx}
\usepackage{subfigure}
\usepackage{url}
\usepackage{color}
\usepackage[linesnumbered,ruled,vlined]{algorithm2e}
\usepackage{listings}
\usepackage{multirow}
\makeindex

\newcommand \Qed {\hfill$\square$}

\newcommand \mpise {MPISE}
\newtheorem{Definition}{ Definition~}
\begin{document}

\title{\mpise: Symbolic Execution of MPI Programs}

\author{Xianjin Fu$^{1,2}$, Zhenbang Chen$^{1,2}$, Yufeng Zhang$^{1,2}$,\\Chun Huang$^{1,2}$,Wei Dong$^{1}$ and Ji Wang$^{1,2}$}

\institute{School of Computer,
    National University of Defense Technology, P. R. China
    \and
        Science and Technology on Parallel and Distributed Processing Laboratory, National University of Defense Technology, P. R. China \\
  \email{\{xianjinfu,zbchen,yfzhang,chunhuang,wdong,wj\}@nudt.edu.cn}}

\maketitle

%
\begin{abstract}
Message Passing Interfaces (MPI) plays an important role in parallel computing. Many parallel applications are implemented as MPI programs. The existing methods of bug detection for MPI programs have the shortage of providing both input and non-determinism coverage, leading to missed bugs. In this paper, we employ symbolic execution to ensure the input coverage, and propose an on-the-fly schedule algorithm to reduce the interleaving explorations for non-determinism coverage, while ensuring the soundness and completeness. We have implemented our approach as a tool, called \mpise{}, which can automatically detect the deadlock and runtime bugs in MPI programs. The results of the experiments on benchmark programs and real world MPI programs indicate that \mpise{} finds bugs effectively and efficiently. In addition, our tool also provides diagnostic information and replay mechanism to help understanding  bugs.
\end{abstract}

%
\section{Introduction}
\label{sec:intro}
In the past decades, Message Passing Interface (MPI) \cite{MPI2.2} has become the \emph{de facto} standard programming model for parallel programs, especially in the filed of high performance computing. A significant part of parallel programs were written using MPI, and many of them are developed in dozens of person-years \cite{Gopalakrishnan:2011:FAM:2043174.2043194}.

Currently, the developers of MPI programs usually use traditional methods to improve the confidence of the programs, such as traditional testing and debugging \cite{GDB}\cite{DDT}. In practice, developers may waste a lot of time in testing, but only a small part of behavior of the program is explored. MPI programs have the common features of concurrent systems, including non-determinism, possibility of deadlock, \emph{etc}. These features make the shortage of testing in coverage guarantee more severe. Usually, an MPI program will be run as several individual processes. The nature of non-determinism makes the result of an MPI program depend on the execution order of the statements in different processes. That is to say, an MPI program may behave differently with a same input on different executions. Hence, sometimes it is harder to find the bugs in an MPI program by a specific program execution.

To improve the reliability of MPI programs, many techniques have been proposed. Basically, we can divide the existing work into two categories: static analysis methods \cite{MPISPIN}\cite{TASS}\cite{Strout:2006:DAM:1156433.1157634} and dynamic analysis methods \cite{Vo:2010:SDD:1884643.1884681} \cite{cav08-isp}. A static method analyzes an MPI program without actually running it. The analysis can be carried out on code level \cite{TASS} or model level \cite{MPISPIN}. Usually, a static method needs to make an abstraction of the MPI program under analysis \cite{MPISPIN}\cite{Strout:2006:DAM:1156433.1157634}. Therefore, many static methods suffer the false alarm problem.

Dynamic methods, such as testing  and runtime verification, need to run the analyzed MPI programs and utilize the runtime information to do correctness checking \cite{KrammerBMR03}\cite{ita}\cite{conf/sc/VetterS00}, online verification \cite{ppopp-VakkalankaSGK08}\cite{Vo:2010:SDD:1884643.1884681}, debugging \cite{DDT}\cite{TotalView}, \emph{etc}. Traditional testing methods work efficiently in practice by checking the correctness of a run under a given test harness. However, testing methods cannot guarantee the coverage on non-determinism even after many runs of a same program input. Other dynamic analysis methods, such as ISP \cite{ppopp-VakkalankaSGK08},  provide the coverage guarantee over the space of non-determinism and scale well, but they are still limited to program inputs. While TASS \cite{TASS} employs symbolic execution and model checking to verify MPI programs, it only works on small programs due to the limited support of runtime library models.

In this paper, we use symbolic execution to reason about all the inputs and try to guarantee the coverage on both input and non-determinism. We symbolically execute the statements in each process of an MPI program to find input-related bugs, especially runtime errors and deadlocks. For the non-determinism brought by the concurrent features, we use an on-the-fly scheduler to reduce the state space to be explored in the analysis, while ensuring the soundness and completeness. Specially, to handle the non-determinism resulted from the wildcard receives in MPI programs, we dynamically match the source of a wildcard receive into all the possible specific sources in a \emph{lazy} style, which avoids the problem of missing bugs. Furthermore, unlike the symbolic execution plus model checking method in \cite{TASS}, which uses an MPI model to simulate the runtime behaviors of MPI library, we use a true MPI library as the model, which enables us to analyze real-world MPI programs.

To summarize, our paper has the following main contributions: firstly, we propose an on-the-fly scheduling algorithm, which can reduce unnecessary interleaving explorations while ensuring the soundness and completeness; secondly, when attacking the non-determinism caused by wildcard receives, we propose a technique, called lazy matching, to avoid blindly matching each process as the source of a wildcard receive, which may lead to false positives; finally, we have implemented our approach in a tool called \mpise{}, and conducted extensive experiments to justify its effectiveness and efficiency in finding bugs in MPI programs.

The rest of this paper is organized as follows. Section 2 introduces the back- ground and shows the basic idea of \mpise{} by motivating examples. Section 3 describes the details of the algorithms implemented in \mpise{}. Section 4 explains our implementation based on Cloud9 and shows the experimental results. Finally, Sections 5 discusses the related work and the conclusion is drawn in Section 6.

\section{Background and Motivating Example}
In this section, we briefly describe symbolic execution and the scope of the MPI APIs we are concerned with, then show how our algorithm works by motivating examples.
\subsection{Symbolic execution}
Symbolic execution \cite{K76} is a SAT/SMT based program analysis technique originally introduced in the 1970s. With the significant improvement in SAT/SMT techniques and computing power, symbolic execution draws renewed interests recently. The main idea is, rather than using concrete values, symbolic execution uses symbolic values as input values, and keeps tracking the results of numerical operations on symbolic values.
Hence, the result of a program under symbolic execution will be symbolic expressions. Most importantly, symbolic execution uses a constraint of symbolic values, called path condition (PC), to represent a path of a program. At the beginning, the path condition is \emph{true}. When encountering a branch statement, symbolic execution explores both directions of the branch. For exploring one direction, symbolic execution records (\emph{i.e.}, conjunction) the condition $cond$ corresponding to the direction in PC and queries an underlying solver with $PC\wedge cond$ to decide whether this direction is feasible. If the answer is yes, symbolic execution will continue to execute the statements following the direction, and PC is update to be $PC\wedge cond$; otherwise, it means the direction is infeasible, thus symbolic execution backtracks to the branch statement, and starts to explore the other direction. The selection of which direction of a branch to explore first can be random or according to some heuristics. Once symbolic execution reaches the end of a program, the accumulated PC represents the constraints that the inputs need to satisfy to drive the program to the explored path.
Therefore, we can consider symbolic execution as a function that computes a set of PCs for a program. Naturally, we can use the PCs of the program to do automatic test generation \cite{Cadar:2008:KUA:1855741.1855756}\cite{Cadar:2011:SES:1985793.1985995}, bug finding \cite{Cadar:2008:KUA:1855741.1855756}\cite{godefroid2008sage}, verification \cite{xianghuadeng2006ase}, \emph{etc}.

According to the before explanation, symbolic execution is a precise program analysis technique, because each PC represents a real feasible path of the program under analysis. Therefore, when used for bug finding, symbolic execution does not suffer from the false alarm problem, and the  bugs found are real bugs. Whereas, one of the major challenge symbolic execution faces is path space exploration, which is theoretically exponential with the number the branches in the program.


\subsection{MPI Programs}
An MPI program is a sequential program in which some MPI APIs are used. The running of an MPI program usually consists of a number of parallel processes, say $P_0, P_1, ..., P_{n-1}$, that communicate via message passings based on MPI APIs and the supporting platform. The message passing operators we consider in this paper include:
\begin{itemize}
\item Send(dest) -send a message to $P_{dest}$ $(dest=0,\dots,n-1)$, which is the destination process of the Send operation. Note that only synchronous communications are considered in this paper, so this operation blocks until a matching receive has been posted.

\item Recv(src) -receive a message from $P_{src}$ $(src=0,\dots,n-1, ANY)$, which is the source process of the Recv operation. Note that the $src$ can take the wildcard value ``ANY", which means this Recv operation expects messages from any process. Because Send and Recv are synchronous, a Send/Recv that \emph{fails} to match with a corresponding Recv/Send would result in a \emph{deadlock}.

\item Barrier() -synchronization of all processes, which means the statements of any process should not be issued past this barrier until all the processes are synchronized. Therefor, an MPI program is expected to eventually reach such a state that all the processes reach their barrier calls. If this does not hold, there would be a deadlock.
\end{itemize}

The preceding three MPI operations are the most important operations we consider in this paper. Actually, they cover the most frequently used \emph{synchronous} communications in MPI programs.

\subsection{Motivating Examples}
 Usually, an MPI program is fed with inputs to perform a computational task, and the bugs of the program may be input-dependent. On the other side, due to the non-determinism feature, even with same inputs, one may find that bugs occur ``sometimes''.

Consider the MPI program in Fig \ref{moti-example}, if the program runs with an input that is not equal to `a', the three processes will finish normally with two matched \textrm{Send} and \textrm{Recv},
as indicated by Fig \ref{fig:subfig:case1}. However, if the program is fed with the input `a', a deadlock may happen, in case that $Proc_1$ receives a message from $Proc_2$ first by a wildcard receive, and then it waits a message from $Proc_2$ and $Proc_0$ also expects $Proce_1$ to receive a message, as shown in Fig \ref{fig:subfig:case3}. Therefore, tools that do not provide input space coverage would surely fail to detect this bug if the program is not fed with `a'. Even if one is lucky enough to run the program with `a', we may still fail to detect the bug if the wildcard receive is matched with $Proc_0$, \emph{e.g.}, the case in Fig \ref{fig:subfig:case2}.


\begin{figure}[width=.9\textwidth]
\vspace{-2mm}
\begin{lstlisting}
int main(int argc, char **argv) {  	
  int x, y, myrank;
  MPI_Comm comm = MPI_COMM_WORLD;

  MPI_Init(&argc, &argv);
  MPI_Comm_rank(comm, &myrank);
  if (myrank==0) {
    x = 0;
    MPI_Ssend(&x, 1, MPI_INT, 1, 99, comm);
  }
  else if (myrank==1) {
    if(argv[1][0]!= 'a') //  argc is exactly 2
      MPI_Recv(&x, 1, MPI_INT, 0, 99, comm, NULL);
    else
      MPI_Recv(&x,1, MPI_INT, MPI_ANY_SOURCE, 99, comm, NULL);

    MPI_Recv(&y, 1, MPI_INT, 2, 99, comm, NULL);
  } else if (myrank==2){
    x = 20;
    MPI_Ssend(&x, 1, MPI_INT, 1, 99, comm);
  }
  MPI_Finalize();
  return 0;
}
\end{lstlisting}
\vspace{-4mm}
\caption{Example showing the need for both input and non-determinism coverage}\label{moti-example}
\vspace{-2mm}
\end{figure}

Thus, for detecting deadlock bugs, we need both input coverage and non-determinism coverage guarantee. The basic idea of our method is: we employ symbolic execution to cover all possible inputs, and explore all the possible matches of a wildcard receive by matching it to any possible source.

\begin{figure}[htpb]
  \centering
\subfigure[X$\neq$’a’]{
\label{fig:subfig:case1}
\includegraphics[width=0.34\textwidth]{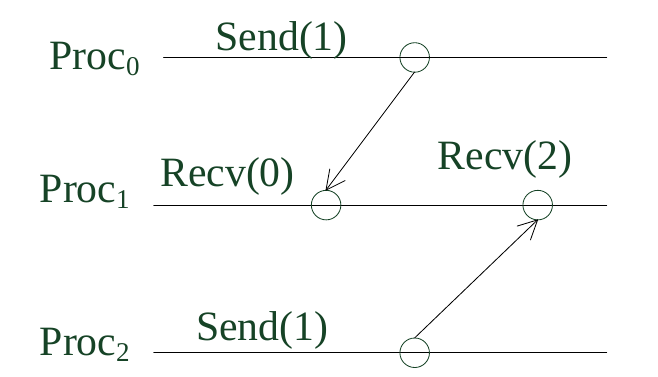}}
\hspace{-0.1em}
\subfigure[X==’a’ and wildcard receive matches with $Proc_0$]{
\label{fig:subfig:case2} 
\includegraphics[width=0.29\textwidth]{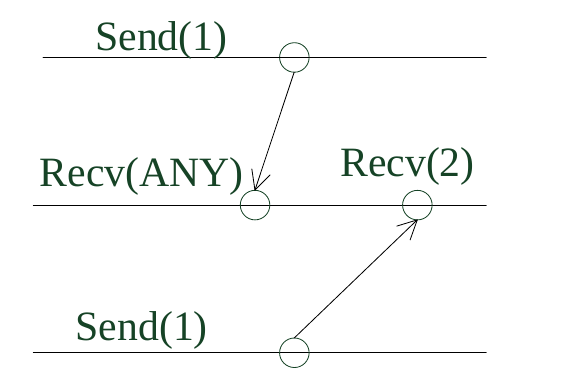}}
\hspace{0.1em}
\subfigure[X==’a’ and wildcard receive matches with $Proc_2$]{
\label{fig:subfig:case3} 
\includegraphics[width=0.29\textwidth]{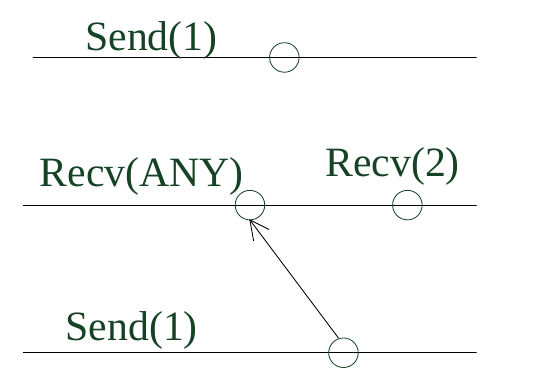}}
\vspace{-1mm}
%
%
%

\caption{Three cases of the program in Fig \ref{moti-example}}\label{fig:3cases}
\vspace{-2mm}
\end{figure}

To be more detailed, since we only symbolically execute one process at a time, we need to decide the exploration order of the processes. Usually, each process of an MPI program has a rank,
 we always start from the smallest ranked process and switch to another process until the current process needs synchronization, such as sending or receiving a message. Thus, the switches during symbolic execution happen \emph{on-the-fly}. Specifically, things become more complex when encountering a Recv(ANY) statement, where we need to delay the selection of the corresponding sending process until all the possible sending statements are encountered.

For the MPI program in Fig \ref{moti-example} run in three processes, we start from $Proc_0$, \emph{i.e.}, the process with rank 0. When executing to line 9, a Send is encountered, which means a synchronization is needed. From the send statement, we know it needs to send a message to $Proc_1$. Thus, we switch to $Proc_1$ and do symbolic execution from the beginning. When the branch statement at line 12 is encountered, and $argv[1][0]$ is symbolic (we suppose it has a symbolic value $X$), the condition $X \neq{}\text{`a'}$ is added to the path condition of the true side and its negation to the false side. We mark here as a backtrack point and has two paths to follow, which are explained as follows:
\begin{description}
  \item[$X\neq{}  \text{`a'}$:] If we explore the true side first, the path condition, \emph{i.e.}, $X \neq \text{`a'}$, is fed to the solver to check the feasibility of the path. Apparently, the solver will answer yes, thus we can continue the symbolic execution of $Proc_1$. Then, Recv(0) is meet and it is exactly matched with the send in $Proc_0$. Therefore, both processes advance, and $Proc_0$ ends while $Proc_1$ goes to Recv(2). In a same manner,  $Proc_1$ gets asleep, we switch to $Proc_2$. Again the two operations matches, the whole execution will end normally, as shown in Fig \ref{fig:subfig:case1}.
  \item[$X =={} \text{`a'}$:] This side is also feasible. The symbolic execution of $Proc_1$ will encounter Recv(ANY), and switches to $Proc_2$. After executing the Send at Line 20, there is no process that can be switched to. All the possible sending processes of the Recv(ANY) in $Proc_1$ are determined. Thus, now we begin to handle the Recv(ANY) by 
      matching it with each possible sending. Suppose we match the Recv(ANY) with the Send of $Proc_0$, we continue to execute $Proc_1$. We encounter another Recv at Line 17 that expects to receive a message from $Proc_2$, then $Proc_1$ and $Proc_2$ advance, and finally the whole execution ends normally, as indicated by Fig \ref{fig:subfig:case2}. On the other hand, if the Recv(ANY) is matched with the Send of $Proc_2$, when encountering the Recv in $Proc_1$, symbolic execution will switch to $Proc_2$, but $Proc_2$ has finished. Then, $Proc_0$ and $Proc_1$ can not terminated. Hence, a deadlock is detected, as shown in Fig \ref{fig:subfig:case3}.
\end{description}

In summary, the deadlock, which may happen in the program in Fig \ref{moti-example} when run in three processes, can only be encountered when the input starts with `a' and the Recv(ANY) in the second process is matched with the Send in the third process. By using our approach, \mpise{} can detect it automatically. The details of our symbolic execution algorithms will be introduced in the next section.


\section{Symbolic execution algorithms}
In this section, we will introduce a general framework for symbolic execution of MPI programs first, and then present a scheduling algorithm during the symbolic execution. Furthermore, to attack the non-determinism brought by wildcard receives, we will present a refined scheduling method, which can ensure the exploration of all the possible matches of a wildcard receive.


%
%


To start with, we introduce some notions first. When symbolically executing a sequential program, the symbolic executor keeps tracking of states, each of which consists of a map that records the symbolic/concrete value of each variable, a program counter and a path condition. For an MPI program, a state during symbolic execution is composed by the states of the parallel processes. A state $s'$ is said to be the successor of a state $s$, if $s'$ can be obtained by symbolically executing a statement in one process. With the notion of \emph{state}, we define a \emph{deadlock} to be the state that has no successor, and at which there is at least one process that does not terminate. Recall that symbolic execution will do state forking when encountering a branch statement. For MPI programs, in addition to branch statements, the concurrency nature can also result in state forking. Theoretically, for the current state, if there are more than one process, say $n$,  that can be executed, there are $n$ possible successor states. Hence, besides the number of branch statements, the number of parallel processes also makes the path space increase exponentially. Algorithm \ref{alg:framework} presents a general framework for symbolic execution of MPI programs.

{\small
\begin{algorithm}[H]
Search($MP$, $n$, $slist$)\{\\
$Active=\{P_0, \dots, P_n\}$ ; $Inactive = \emptyset$\;
NextProcCandidate= -1; worklist = \{initial state\}\;
\While{\emph{(}worklist is not empty\emph{)}}
{
$s$ = pick next state\;
$p$ = \emph{Scheduler}($s$)\;
\If{$p \neq null$}{
$stmt$ = the next statement of $p$\;
SE($s$, $p$, $stmt$);
}
}
\}
\caption{Symbolic Execution Framework}\label{alg:framework}
\end{algorithm}
}

Basically, the symbolic execution procedure is a worklist-based algorithm. The input consists of an MPI program, the number of the parallel running processes and the symbolic variables. At the beginning, only the initial state, \emph{i.e.}, composed by the initial states of all the processes, is contained in the worklist. Then, new states can be derived from the current state and put into the worklist. State exploration is done if there is no state in the worklist. Because of state forking, we usually have a way for space exploration, such as depth first search (DFS) and breadth first search (BFS). Clearly, it is hard or even impossible to explore the whole path space. In fact, for the state forking introduced by the concurrent feature, sometimes there is no need to add all the possible successor states to the worklist, which can still capture the behavior of the program precisely in our context. Hence, different from the usual symbolic execution algorithm, in our algorithm, we first select a state from worklist (Line 5, where a search algorithm can be used), then we make a decision (Line 6, the details of which will be given in Section \ref{sub:sched}) of which process is scheduled for symbolic execution. Finally, we symbolically execute the next statement of the scheduled process, in which some new states may be generated.

Basically, for the non-communication statements in an MPI program, the symbolic execution semantics is same as usual. In the following of this section, we will concentrate on explaining the scheduling of the processes and the handling of the communication operations.
\vspace{-4mm}
\subsection{On-the-fly scheduling}
\label{sub:sched}

With the general framework in Algorithm \ref{alg:framework}, we introduce our scheduler here, aiming to avoid naively exploring the interleavings of all the processes.
For each process of an MPI program during symbolic execution, the process is \emph{active} if it is not asleep. Usually, we make a process asleep when the process needs to communicate but the corresponding process is not ready, whose details will be given in Algorithm \ref{alg:se}. We maintain the current status of each process via two sets: $Active$ and $Inactive$. At beginning, all the processes are contained in $Active$. If a process is made to be asleep, it will be removed from $Active$ and added to $Inactive$. Because we schedule the processes on-the-fly, we use a global variable $NextProcCandidate$ to denote the index of the next process to symbolically execute. The following \mbox{Algorithm \ref{alg:scheduler}} gives how to do scheduling.

\vspace{2mm}
{\small
\begin{algorithm}[H]
Scheduler(s)\{\\
\If{$NextProcCandidate != -1$ and $Proc_{NextProcCandidate}$ is active}
{
$Next = NextProcCandidate$\;
$NextProcCandidate = -1$\;
\Return $Proc_{Next}$\;
}
\ElseIf{$Active \neq \emptyset$}{
\Return the process $p'$ with the smallest rank in $Active$\;
}
\If {$Inactive \neq \emptyset $}  {Report Deadlock\; }
\}
\caption{Scheduling the Next Process for Symbolic Execution}\label{alg:scheduler}
\end{algorithm}
}\vspace{2mm}

First, we check whether there is a next process that needs to be executed and is also active. If there exists one, the process identified by $NextProcCandidate$ will be selected, and the next process global variable is reset (Line 1$\sim$5); otherwise, we return the active process with the smallest rank if exists (Line 6$\sim$7). Finally, if there is no active process that can be scheduled, and the $Inactive$ set is non-empty, \emph{i.e.}, there exists at least one process that does not terminate, we report that a deadlock is found (Line 8$\sim$9).

Now, we explain how to symbolically execute each statement. In \mbox{Algorithm \ref{alg:se}}, we mainly give the handling for MPI APIs considered in this paper. The local statements in each process do not influence the other processes, and the symbolic execution of basic statements, such as assignment and branch, is the same with the traditional approach \cite{Cadar:2008:KUA:1855741.1855756}. Hence, the symbolic execution of local statements is omitted for the sake of space. In Algorithm \ref{alg:se}, Advance($S$) denotes the procedure in which the program counter of each process in $S$ will be advanced, and \mbox{Match($p$, $q$)} denotes the procedure in which the synchronization between $p$ and $q$ happens, \emph{i.e.}, the receiver receives the data sent by the sender, and the program counters of $p$ and $q$ will be both advanced.

If a $Send(dest)$ is encountered and there is a process in $Inactive$ that matches the statement, we move that process from $Inactive$ to $Active$ (Line 5) and advance the two processes (Line 6). If there is no process that can receive the message, we add this process into \emph{Inactive} set (Line 8), and switch to the destination process of the send operation (Line 9). The execution of a receive operation is similar, except that when the receive operation is a wildcard receive, we make the current process asleep (the reason will be explained in Section \ref{section:wr}).

{\small
\begin{algorithm}[H]
SE($s$, $p$, $stmt$)\{\\
\Switch{kindof\emph{(}$stmt$\emph{)}}{
\Case{Send\emph{(}dest\emph{)}}{
\If{$stmt$ has a matched process $q \in Inactive$}{
$Inactive = Inactive \setminus \{q\}$;\ $Active = Active \cup \{q\}$\;
Match($p$, $q$)\;
}
\Else{
$Inactive = Inactive \cup \{p\}$\;
$NextProcCandidate= dest$\;
}
\Return\;
}
\Case{Recv\emph{(}src\emph{)}}{
\If{src \emph{!= MPI\_ANY\_SOURCE} }{
\If{$stmt$ has a matched process $q \in Inactive$}{
$Inactive = Inactive \setminus \{q\}$;\ $Active = Active \cup \{q\}$\;
Match($p$, $q$)\;
}
\Else{
$Inactive = Inactive \cup \{p\}$\;
$NextProcCandidate = src$\;
}
}
\Else{
$Inactive = Inactive \cup \{p\}$\;
}
\Return\;
}
\Case{Barrier}{
\If{$mc_b == \emptyset$}{
$mc_b = \{P_0,\dots, P_n\} \setminus \{ p \}$\;
$Inactive = Inactive \cup \{p\}$\;
}
\Else{
$mc_b = mc_b \setminus \{p\}$\;
\If{$mc_b == \emptyset$}{
Advance($\{P_0, \dots, P_n\}$)\;
$Inactive = \emptyset$;\ $Active = \{P_0, \dots, P_n\}$\;
}
\Else{
$Inactive = Inactive \cup \{p\}$\;
}
}
\Return\;
}
\Case{Exit}
{
$Active = Active \setminus \{p\}$\;
\Return\;
}
}
Advance($\{p\}$)\;
\}
\caption{Symbolic Execution of a Statement}\label{alg:se}

\end{algorithm}
}

For handling barriers, we use a global variable $mc_b$ to denote the rest processes that need to reach a barrier for a synchronization. When a barrier statement is encountered, if $mc_b$ is empty, we initialize $mc_b$ to be the set containing the rest processes (Line 24) and add the current process into $Inactive$ (\mbox{Line 25}). If $mc_b$ is not empty, we remove the current process from $mc_b$. Then, if $mc_b$ is empty, \emph{i.e.}, all the processes have reached a barrier, we can advance all the processes (Line 29) and make all the processes active (Line 30); otherwise, we add the current process into \emph{Inactive} set (Line 32). When encountering an Exit statement, which means the current process terminates, we remove the current process from $Active$ (Line 35). 

In summary, according to the two algorithms, the symbolic execution process will continue to execute the active process with the smallest rank until a preemption happens caused by an unmatched MPI operation. From a state in symbolic execution, we do not put all the possible states into the worklist, but only the states generated by the current process.   This is the reason why we call it on-the-fly scheduling. Actually, we only explore a sub space of the whole program path space, but without sacrificing the ability of finding deadlock bugs. The correctness of our on-the-fly scheduling algorithms is guaranteed by the following theorem, whose proof is given in appendix.


\begin{theorem}\label{theorem-1}
Given a path of an MPI program from the initial state to a deadlocked state, there exists a path from the initial state to the same deadlocked state obtained by the on-the-fly scheduling. And vice versa.
\end{theorem}

\subsection{Lazy matching algorithm}\label{section:wr}

\begin{figure}[!htp]
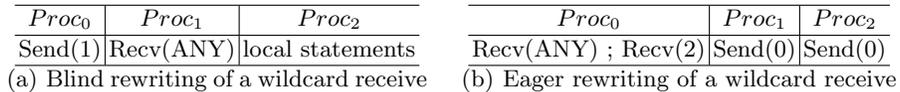

\vspace{-2mm}
  \centering
\subfigure[Blind rewriting of a wildcard receive]{
 \label{fig:brewrite}
 \begin{tabular}{c|c|c}
 \hline
  $Proc_0$          & $Proc_1$            &     $Proc_2$\\
 \hline
    Send(1)        & Recv(ANY)  &    local statements\\
 \hline
 \end{tabular}
 }
 \quad
\subfigure[Eager rewriting of a wildcard receive]{
 \label{fig:erewrite}
 \begin{tabular}{c|c|c}
 \hline
  $Proc_0$          & $Proc_1$            &     $Proc_2$\\
 \hline
  Recv(ANY) ; Recv(2)      & Send(0)   &    Send(0) \\
 \hline
 \end{tabular}
 }
 \vspace{-2mm}
\caption{Rewriting of a wildcard statement}
 \vspace{-2mm}
\end{figure}

Note that so far, we do not treat wildcard receives. Actually, wildcard receives are one of the major reasons of non-determinism. Clearly, we cannot blindly rewrite a wildcard receive. For example, in Fig \ref{fig:brewrite}, if we force the wildcard receive in $Proc_1$ to receive from $Proc_2$, a deadlock will be reported, which actually will not happen. 
In addition, if we rewrite a wildcard receive immediately when we find a possible match, we still may miss bugs. As shown in Fig \ref{fig:erewrite}, if we match the wildcard receive in $Proc_0$ with the send in $Proc_1$, the whole symbolic execution will terminate successfully, thus a deadlock, which will appear when the wildcard receive is matched with the send in $Proc_2$, is missed. 

To solve this problem, we employ a lazy style approach instead of an eager one. That is, we delay the selection of the send candidate of a wildcard receive until the whole symbolic execution procedure blocks. To be detailed, when the symbolic execution encounters a wildcard receive, we would make the current process asleep (Line 20 in Algorithm \ref{alg:se}), waiting for all possible senders. When a matched send is found, the current process will also be made asleep, and we switch to the next active process. When there is no process that can be scheduled, \emph{i.e.}, all the processes are in $Inactive$, we match the wildcard receive to each possible matched send by forking a successor state for each one. Thus, Algorithm \ref{alg:scheduler} needs to be refined to handle wildcard receives. The refined parts are given as follows.

{
\small{
\begin{algorithm}[H]
Scheduler($s$)\{\\
...\\
\If {$Inactive \neq \emptyset $}  {
\If {Exists a \emph{Recv(ANY)} process in $Inactive$} {
$PS = Inactive$\;
\For{each \emph{Recv(ANY)} process $p \in Inactive$}{
\For{each matched process $q \in Inactive$ of $p$}{
$Inactive = PS \setminus \{p, q\}$;\ $Active = \{p, q\}$\;
AddState($s$, $p$, $q$)\;
}
}
\Return null\;
}
\Else{
Report Deadlock\;
}
}
\}
\caption{Refined Scheduling for Handling Wildcard Receives}\label{alg:refined-scheduler}
\end{algorithm}
}
}
\vspace{2mm}

For each process encountering a wildcard receive in $Inactive$, we add a new state for each of its matched sender processes (Line 9). The AddState($s$, $p$, $q$) denotes a procedure that does the synchronization between $p$ and $q$, advances both $p$ and $q$, and adds the new state to the worklist. Thus, we are exploring all the possible cases of a wildcard receive. If there are multiple Recv(ANY) processes, we are interleaving the matches of all the processes. The example in Fig \ref{fig:breakup} demonstrates this situation. When all the processes are asleep, if we match the Recv(ANY) in $Proc_1$ with the send in $Proc_0$ first, no deadlock will be detected; otherwise, if we match the Recv(ANY) in $Proc_2$ with the send in $Proc_3$ first, a deadlock will be detected.
\vspace{-4mm}
\begin{figure}[!htpb]
  \centering
 \begin{tabular}{l|l|l|l}
 \hline
  $Proc_0$          & $Proc_1$            &     $Proc_2$ & $Proc_3$\\
  \hline
    Send(to:1) \qquad       & Recv(from:ANY)\qquad &   Recv(from:ANY)  \qquad &Send(to:2)\\
  \hline
  & Recv(from:3)\qquad& &Send(to:1)\\
  \hline
     \end{tabular}
  \caption{Multiple wildcard receives}\label{fig:breakup}
  \vspace{-3mm}
\end{figure}

Therefore, after considering wildcard receives, the matches of different wildcard receives are not independent. We deals with this problem by naively interleaving the match orders of wildcard receives. This leads to redundant interleavings, but dose not miss interleaving-specific deadlocks. The optimization is left to our future work. The proof of correctness of our handling for wildcard receives 
is provided in appendix.



\section{Implementation and Experiments}
\label{Sec: Experiments}

\subsection{Implementation}
\begin{figure}[!htpb]
\vspace{-2mm}
  \centering
  \includegraphics[width=0.8\textwidth]{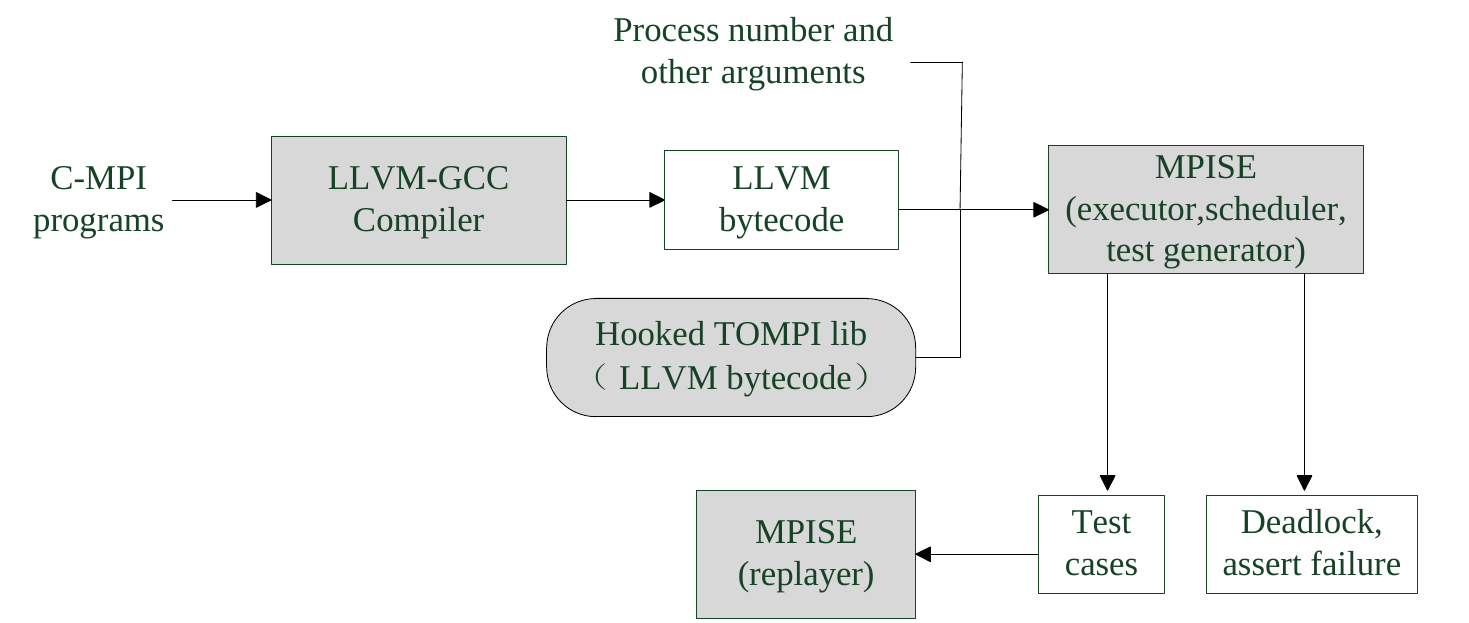}\\
  \caption{The architecture of \mpise{}. }\label{fig:arch}
  \vspace{-3mm}
\end{figure}

We have implemented our approach as a tool, called \mpise{}, based on Cloud9 \cite{Bucur:2011:PSE:1966445.1966463}, which is a distributed symbolic executor for C programs. Cloud9 enhances KLEE \cite{Cadar:2008:KUA:1855741.1855756} by enabling the support of most POSIX interfaces and parallelism. The architecture of \mpise{} is shown in Fig \ref{fig:arch}.
\vspace{-2mm}

The target MPI programs written in C is fed into LLVM-GCC compiler to obtain the LLVM bytecode, which will be linked with a pre-compiled library, \emph{i.e.}, TOMPI \cite{Demaine97athreads-only}, as well as the POSIX runtime library. Then, the linked executable program will be symbolically executed. Basically, TOMPI is a platform that uses multi-threads to simulate the running of an MPI program. TOMPI provides a subset of MPI interfaces, which contains all the MPI APIs we consider in this paper. An MPI program can be compiled and linked with TOMPI libraries to generate a multi-thread executable, which is supposed to generate the same output as that of the parallel running of the MPI program. Hence, we use TOMPI 
as the underlying MPI library. By using TOMPI, we can use the support for concurrency in Cloud9 to explore the path space of an MPI program run with a specific number of processes. When a path ends or a deadlock is detected, \mpise{} records all the information of the path, including the input, the orders of message passings, \emph{etc}. For each path, we generate a corresponding test case, based on which one can use \emph{replayer} to reproduce a concrete path.
Compared with Cloud9, our implementation of \mpise{} consists of the following new features:

\begin{itemize}
\item \textbf{New scheduler.} Cloud9 employs a none-preemptive scheduler, \emph{i.e.}, a process would keep being executed until it gives up, such as encountering an explicit preemption call or process exit.
Clearly, we need a new scheduler for \mpise{}. We have implemented our on-the-fly scheduler that can schedule the MPI processes according to the algorithms in Sections \ref{sub:sched} \& \ref{section:wr}.
\item \textbf{Environment support for MPI APIs.} Cloud9 does not ``recognize" MPI operations, while \mpise{} makes the symbolic engine know MPI operations based on TOMPI, including MPI\_Send, MPI\_Ssend, MPI\_Recv, MPI\_Barrier, \emph{etc}. The message passing APIs are dealt specially for scheduling, while other MPI APIs are treated as normal function callings.
\item \textbf{Enhanced Replay.} \mpise{} can replay each generated test case of an MPI program, which can help user to diagnosis bugs such as deadlock and assertion failure. The replayer of \mpise{} extends the replayer component of Cloud9 by using the on-the-fly schedule when replaying a test case. During replaying, the replayer uses the recorded input to feed the program, and follows the recorded schedules to schedule the processes.
\item \textbf{Enhanced POSIX model.} \mpise{} heavily depends on the library models it uses. However, the POSIX model provided by Cloud9 is not sufficient for us to symbolically execute MPI programs. The reason is we need to maintain a process specific data area for each process when symbolically executing each process. Because we use multi-thread programs to simulate the behaviour of MPI programs, we have improved the mechanism for multi-thread programs in Cloud9 to support maintaining thread specific data.
\end{itemize}

\subsection{Experimental evaluation}\label{subSec:ee}

We have conducted extensive experiments to validate the effectiveness and scalability of \mpise{}. All the experiments were conducted on a Linux server with 32 cores and 250 GB memory.

Using \mpise{} to analyze the programs in the Umpire test suite \cite{ppopp-VakkalankaSGK08},
we have successfully analyzed 33 programs, \emph{i.e.}, either no deadlock detected or detecting a deadlock as expected.

The Umpire test case are input-independent, \emph{i.e.} the inputs have nothing to do with whether a deadlock happens. Hence, we conduct the experiments on the programs with input-dependent deadlocks. The conducted test cases mainly cover two typical situations of deadlock \cite{CPE:CPE701}: point-to-point ones and collective ones. Point-to-point deadlocks are usually caused by (1). a send/receive routine has no corresponding receive/send routine; (2). a send-receive cycle may exist due to the improper usage of send and receive. Collective deadlocks are typically caused by (1). missed collective routines (such as Barrier); (2). improper ordering of some point-to-point or/and collective routines.

In our experiments, we also use ISP and TASS to analyze the programs. Fig \ref{fig:er} displays the experimental results, including those of ISP and TASS.

\begin{figure}[!htpb]
  \centering
{\small
 \begin{tabular}{|c|c|c|c|c|c|c|c|}
 \hline
 &Program & \multicolumn{2}{|c|}{ISP} & \multicolumn{2}{|c|}{TASS} & \multicolumn{2}{|c|}{\mpise{}}\\
 \cline{3-8}
 & & Result & Time(s) & Result & Time(s) & Result & Time(s)\\
 \hline
 & anysrc-deadlock.c & Deadlock & 0.126 & Fail & 1.299 & Deadlock & 1.59\\
 \cline{2-8}
 & basic-deadlock.c & Deadlock & 0.022 & Fail & 1.227 & Deadlock & 1.46\\
 \cline{2-8}
 Input& collect-misorder.c & Deadlock & 0.022 & Fail & 0.424 & Deadlock & 1.48\\
 \cline{2-8}
 Indep-& waitall-deadlock.c & Deadlock & 0.024 & Fail & 1.349 & Deadlock & 1.49\\
 \cline{2-8}
 endent& bcast-deadlock.c & Deadlock & 0.021 & Fail & 0.493 & Deadlock & 1.40\\
 \cline{2-8}
 & complex-deadlock.c & Deadlock & 0.023 & Fail & 1.323 & Deadlock & 1.46\\
 \cline{2-8}
 & waitall-deadlock2.c & Deadlock & 0.024 & Fail & 1.349 & Deadlock & 1.48\\
 \hline
 & barrier-deadlock.c & No & 0.061 & Fail & 0.863 & Deadlock & 1.71\\
 \cline{2-8}
 & head-to-head.c & No & 0.022 & Fail & 1.542 & Deadlock & 1.67\\
 \cline{2-8}
 Input& rr-deadlock.c & No & 0.022 & Fail & 1.244 & Deadlock & 1.67\\
 \cline{2-8}
 Depe-& recv-any-deadlock.c & No & 0.022 & Deadlock & 1.705 & Deadlock & 1.70\\
 \cline{2-8}
 ndent&cond-bcast.c&No&0.021&No&1.410&Deadlock&1.63\\
 \cline{2-8}
 &collect-misorder.c&No&0.023&Deadlock&1.682&Deadlock&1.85\\
 \cline{2-8}
 &waitall-deadlock3.c&No&0.104&Fail&1.314&Deadlock&1.78\\
 \hline
 \end{tabular}
}
\caption{Experimental results}\label{fig:er}
\vspace{-4mm}
\end{figure}
In Fig \ref{fig:er}, we divide the experimental results into two categories: input independent programs and input dependent ones. For each category, we select programs that can deadlock caused by different reasons, including  head to head receive, wait all, receive any, \emph{etc}. For each input dependent program, we generate the input randomly when analyzing the program with ISP, and analyze the program for 10 times, expecting to detect a deadlock. The execution time of analyzing each input dependent program with ISP is the average time of the 10 times of runnings. According to the experimental results, we can conclude as follows:

\mpise{} can detect the deadlock in all the programs. ISP misses the deadlock for all the input dependent programs. TASS fails to analyze most of programs. Thus, \mpise{} outperforms ISP and TASS for all the programs in Fig \ref{fig:er}. The reason is, \mpise{} uses symbolic execution to have an input coverage guarantee, and the scheduling algorithms ensures that any deadlock caused by the MPI operations considered in this paper will not be missed. In addition, we utilize TOMPI and Cloud9 to provide a better environment support for analyzing MPI programs. The reason of the common failure of TASS is that TASS does not support many  APIs, such as \emph{fflush}(stdout) of POSIX and MPI\_Get\_Processor\_Name of MPI, and needs manually modifying the analyzed programs.

For each program, the analysis time using \mpise{} is longer than that of using ISP or TASS. The reason is two fold: firstly, we need to symbolically execute the bytecodes including those of the underlying MPI library, \emph{i.e.}, TOMPI. For example, for the input dependent program cond-barrier-deadlock.c, the number of the executed instructions is 302625.  Secondly, the time used by \mpise{} includes the linking time of the target program byte code and TOMPI library.  In addition, we need to record states and do solving during symbolic execution, which also needs more time than dynamic analysis.

For the rest programs in Umpire test suite, \mpise{} either reports a deadlock that actually does not exist or aborts during symbolic execution. The reason is we only consider the synchronous communications in MPI programs, or some advanced MPI operations, such as MPI\_Type\_vector, are not supported by \mpise{}. The improvement with respect to these aspects is our future work.


\begin{figure}[!h]
\vspace{-2mm}
  \centering
\subfigure[Instruction count of CPI]{
\label{fig:cpi:ins}
\includegraphics[width=0.45\textwidth]{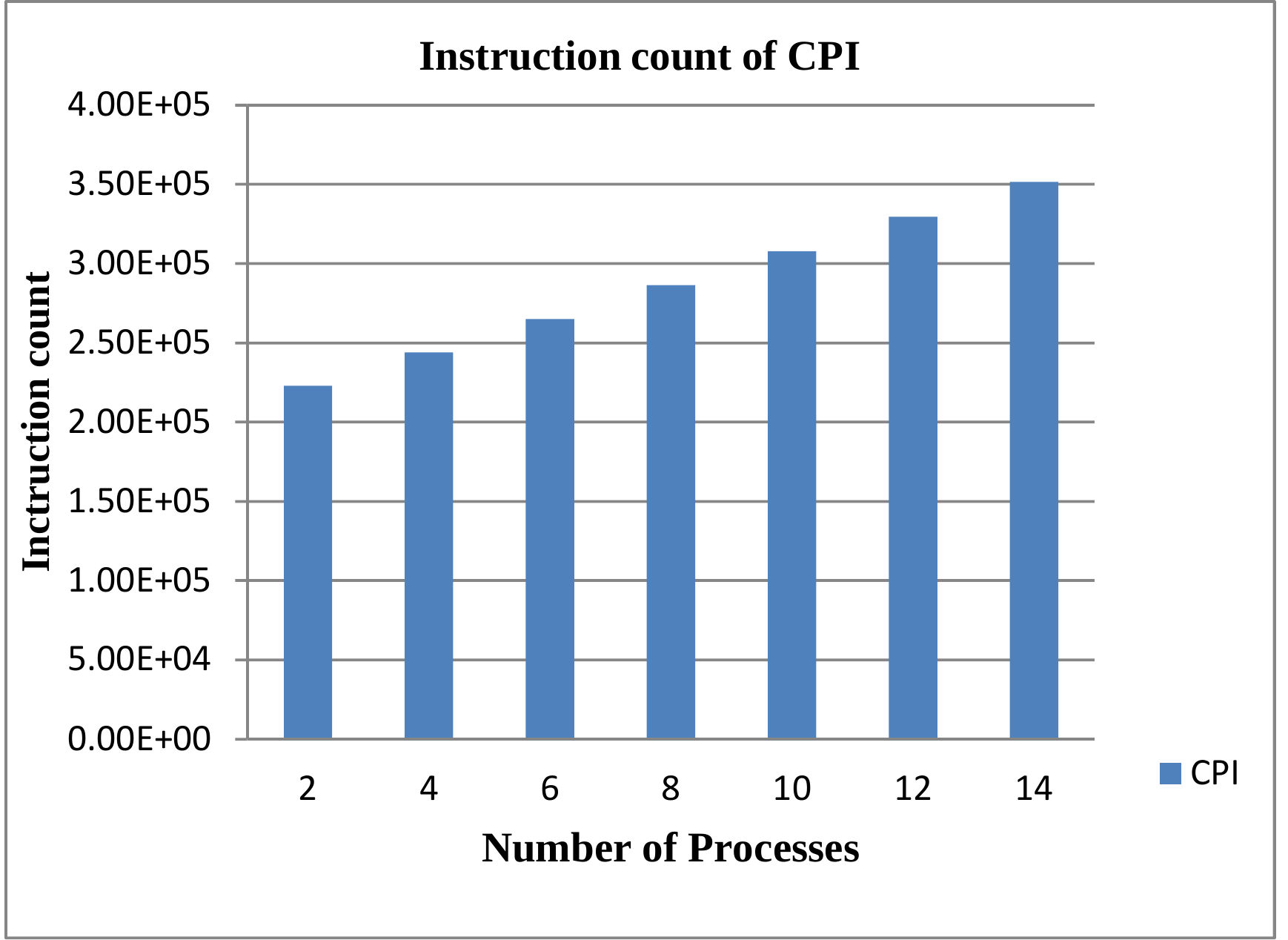}}
\hspace{-0.1em}
\subfigure[Symbolic execution time of CPI]{
\label{fig:cpi:time} 
\includegraphics[width=0.4\textwidth]{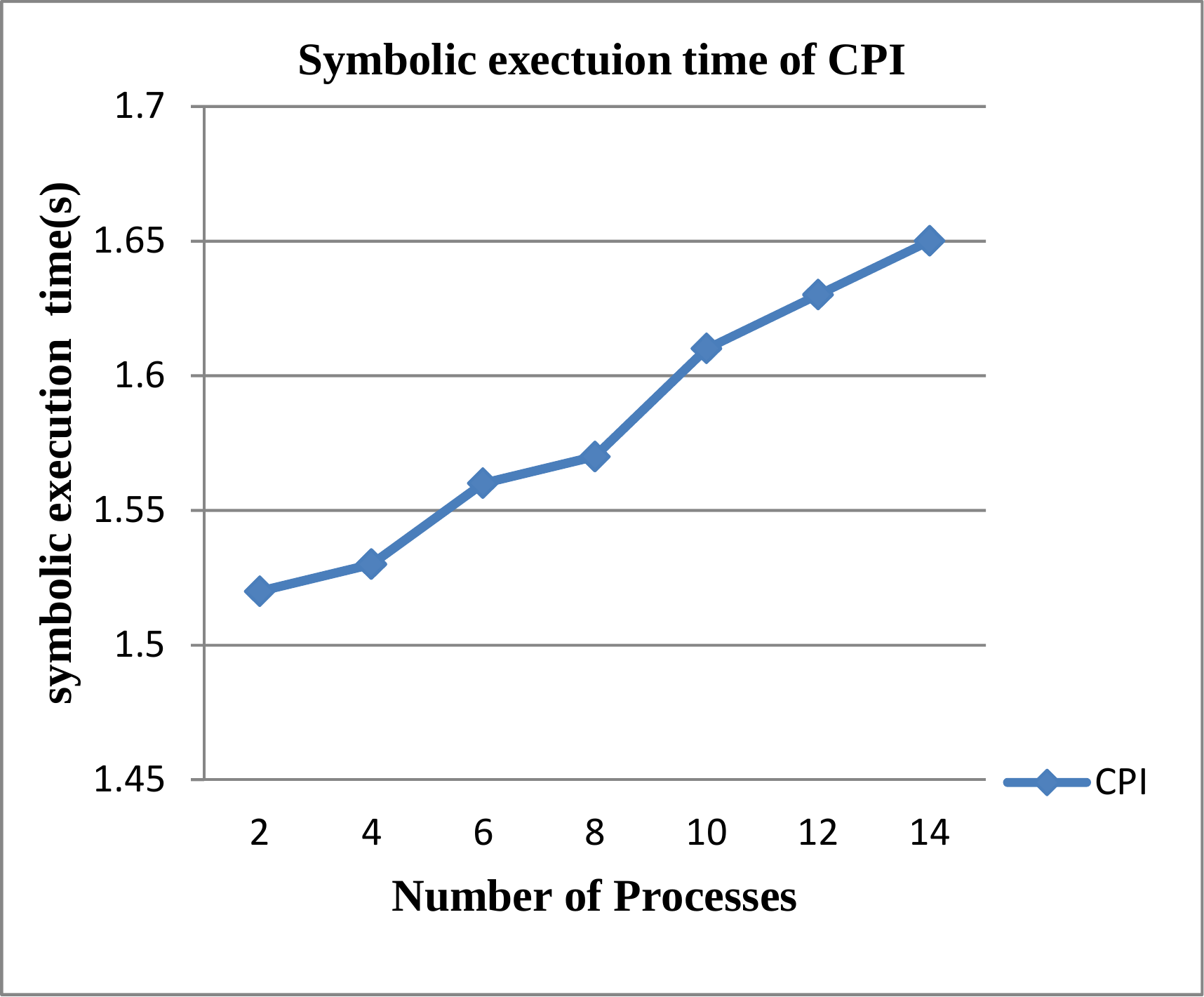}}
  \centering
\subfigure[Instruction count of DT]{
\label{fig:dt:ins}
\includegraphics[width=0.45\textwidth]{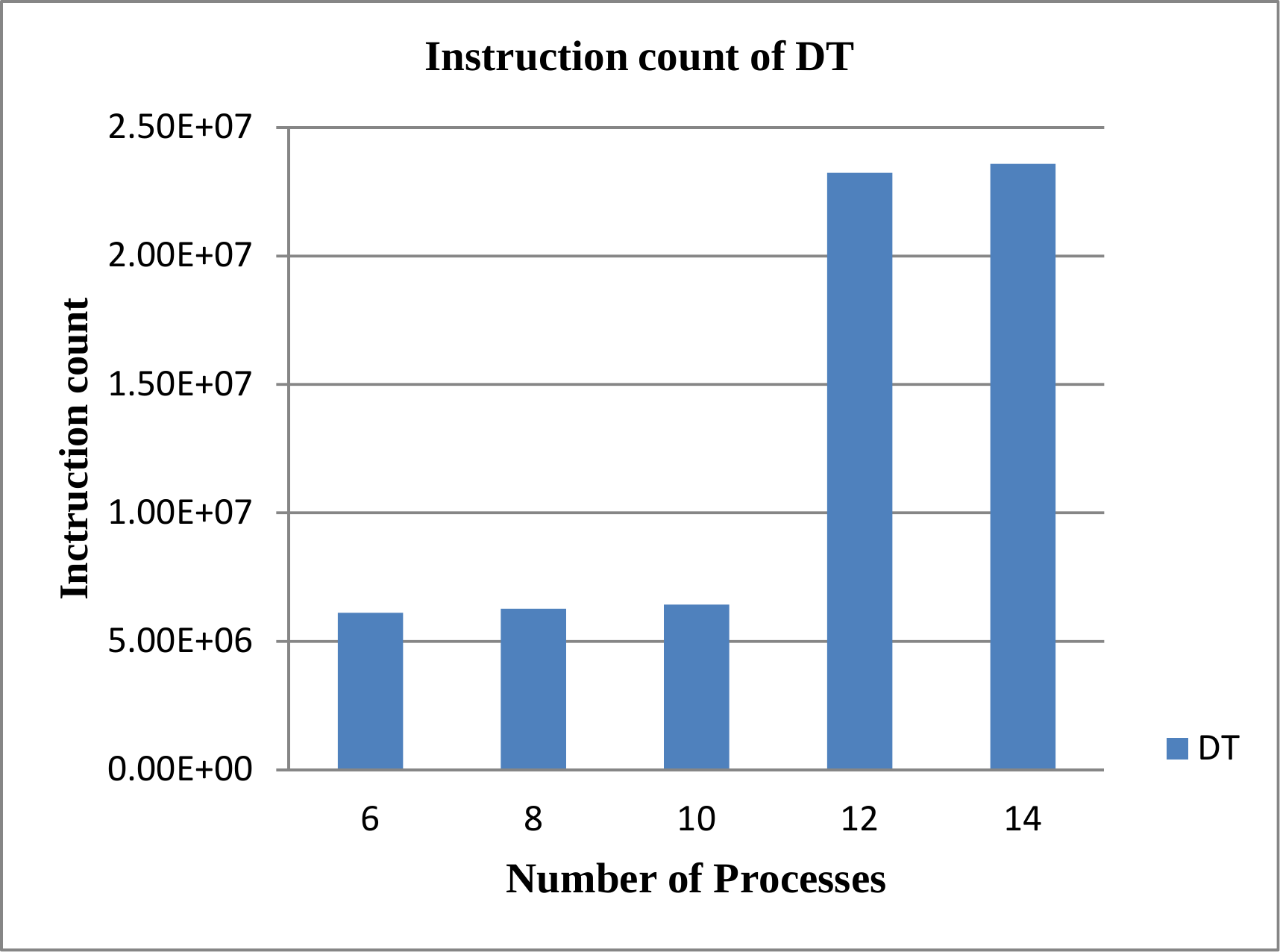}}
\hspace{-0.1em}
\subfigure[Symbolic execution time of DT]{
\label{fig:dt:time} 
\includegraphics[width=0.4\textwidth]{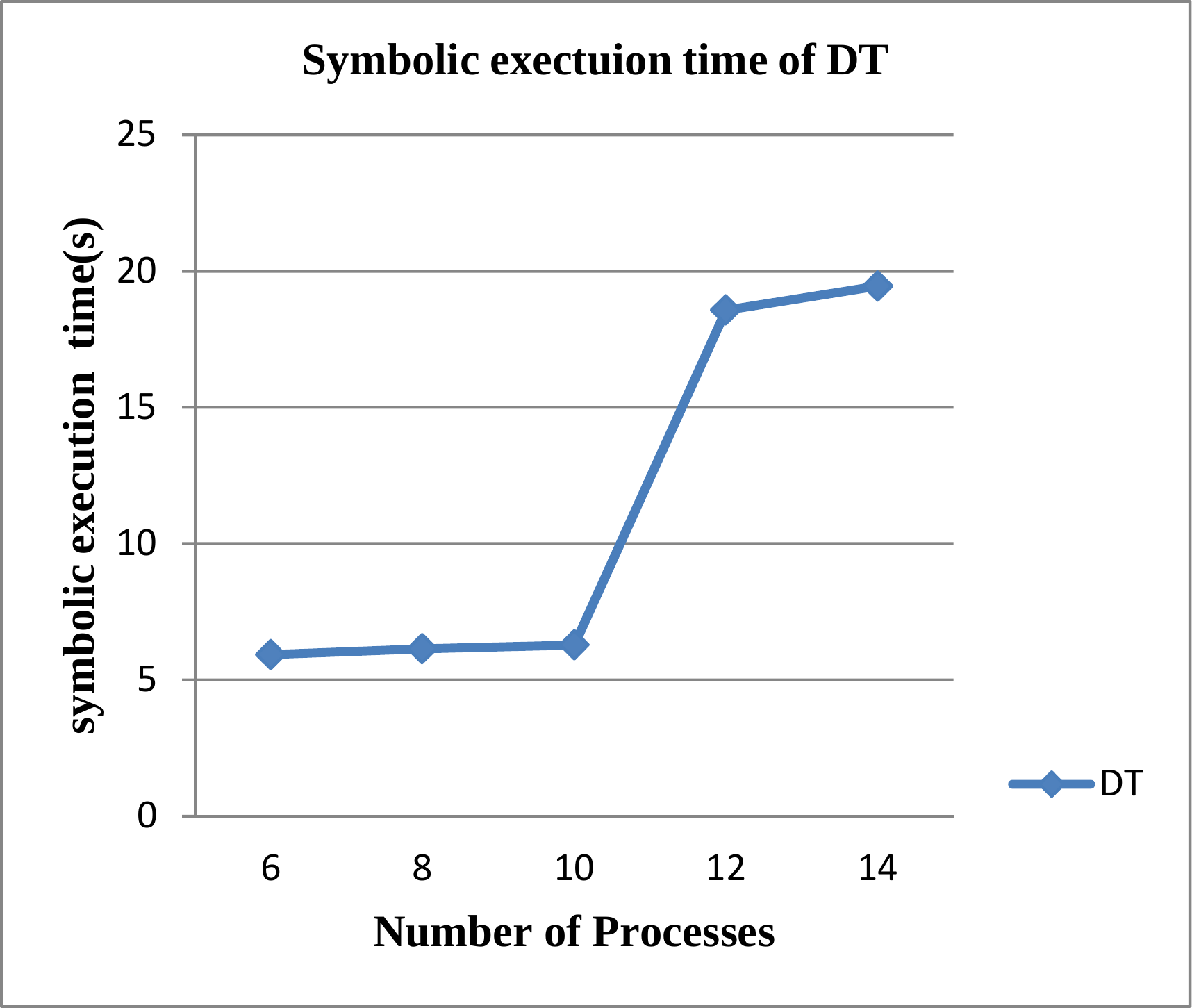}}
\subfigure[Instruction count of IS]{
\label{fig:is:ins}
\includegraphics[width=0.45\textwidth]{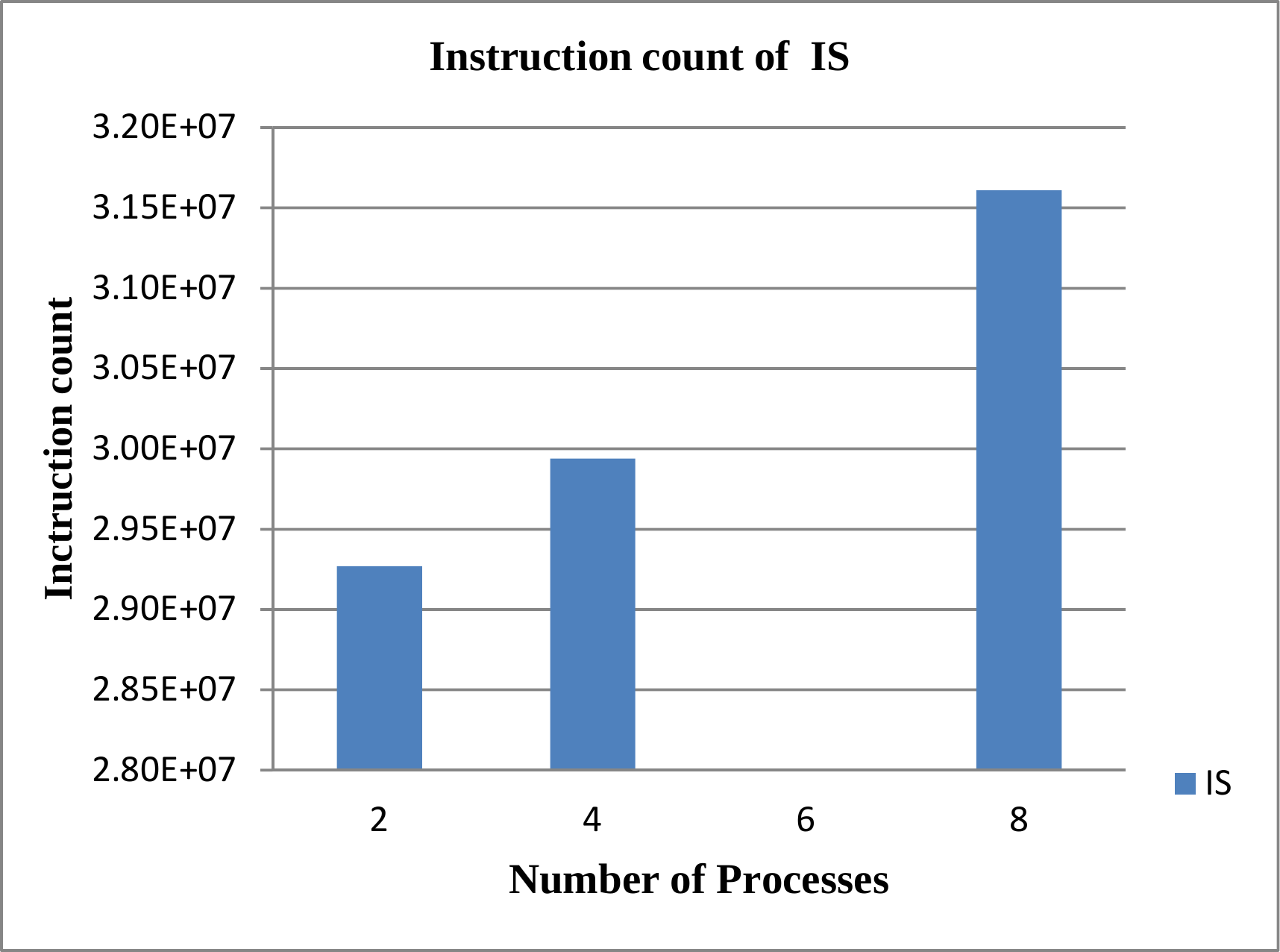}}
\hspace{-0.1em}
\subfigure[Symbolic execution time of IS]{
\label{fig:is:time} 
\includegraphics[width=0.4\textwidth]{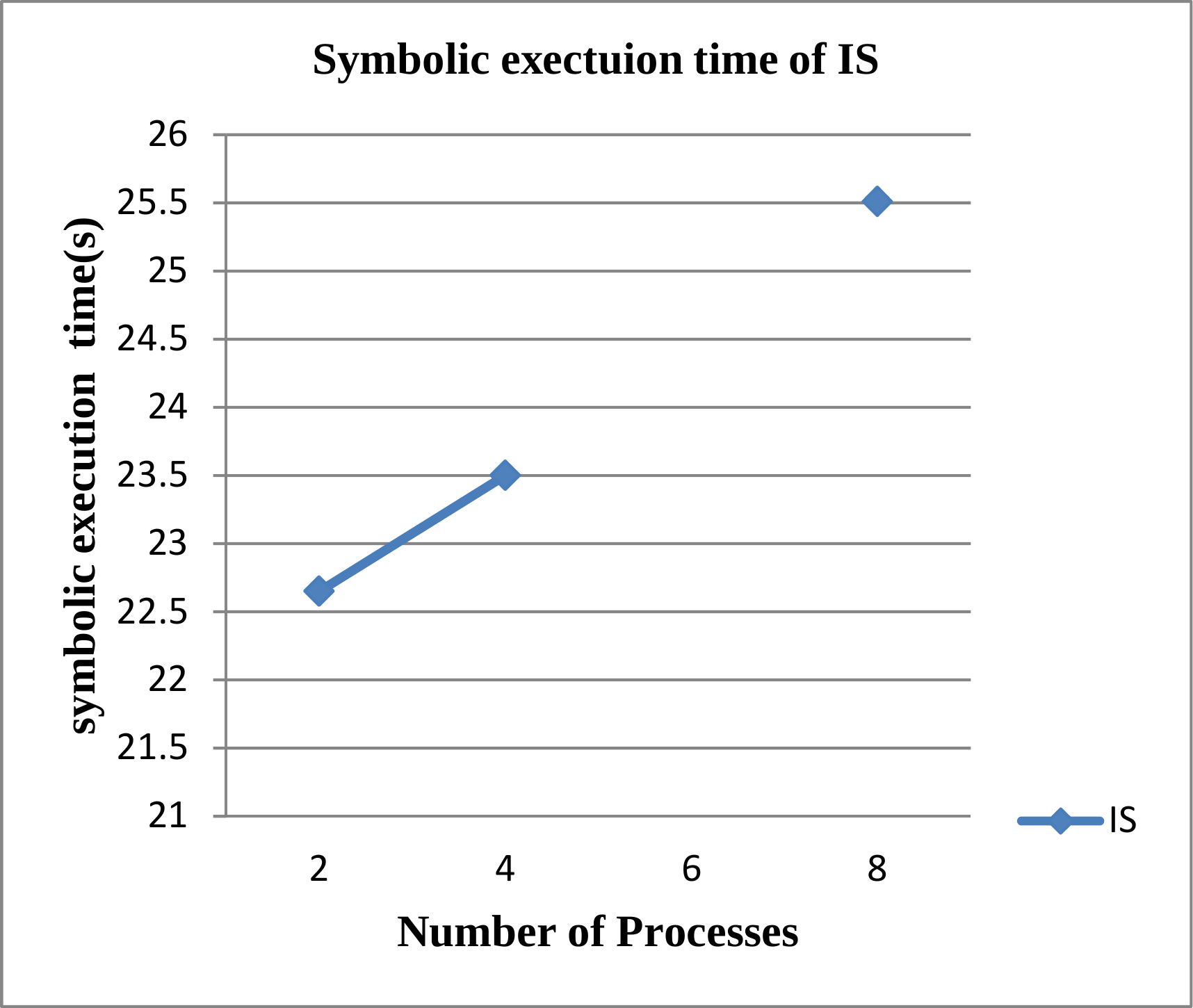}}
\caption{The experimental results under different numbers of processes}
\label{fig:realworld}
\vspace{-5mm}
\end{figure}

To validate the scalability of \mpise{}, we use \mpise{} to analyze three real world MPI programs, including an MPI program (CPI) for calculating $\pi$ and two C MPI programs (DT and IS) from NSA Parallel Benchmarks (NPB) 3.3 \cite{NPB} with class S. The LOC of DT is 1.2K, and the program needs an input that is either BH, WH or SH for showing the communication graph name. IS is an MPI program for integer sorting, and the LOC of IS is 1.4K. \mpise{} can analyze these three programs successfully, and no deadlock is found. We make the input symbolic and symbolically execute all the three MPI programs under different numbers of parallel processes. The experimental results are displayed Fig \ref{fig:realworld}. Because IS can only be run with $2^n$ ($n \ge 1$) processes, we do not have results for the case of 6 processes.

From Fig \ref{fig:realworld}, we can observe that, for all the three programs, the number of the executed instructions and the symbolic execution time do not increase exponentially with respect to the number of processes. It justifies that \mpise{} avoids the exponential increasing of instructions or symbolic execution time caused by the parallelism by the on-the-fly scheduling algorithms. Note that we make the input of DT symbolic ones, and this program aborts early when fed with input BH and the process number that is less than 12,  this explains the sudden rise of both analyze time and instructions when the number of processes goes from 10 to 12 in Fig.\ref{fig:dt:ins} and Fig.\ref{fig:dt:time}.

%
\section{Related Work}
\label{Sec: Related}

There are already some existing work for improving the reliability of MPI programs \cite{Gopalakrishnan:2011:FAM:2043174.2043194}. Generally, they often fall into one of the following two categories: debugging and testing methods, and verification methods.

Debugging and testing tools often scale well, but depend on concrete inputs to run MPI programs, expecting to find or locate bugs. Debugging tools such as TotalView \cite{TotalView} and DDT \cite{DDT} are often effective when the bugs can be replayed consistently. Whereas, for MPI programs, reproducing a concurrency bug caused by non-determinism is itself a challenging problem. Another kind of tools, such as Marmot \cite{KrammerBMR03}, the Intel Trace Analyzer and Collector \cite{ita} and MUST \cite{Hilbrich:2012:MRE:2388996.2389037}, intercept MPI calls at runtime and record the running information of an MPI program, and check runtime errors, deadlock or analyze performance bottlenecks based on the recorded runtime information. These tools often need to recompile or relink MPI programs, and also depend on the inputs and the scheduling of each running.

Another line of tools are verification tools. Dynamic verification tools, such as ISP \cite{ppopp-VakkalankaSGK08} and DAMPI \cite{Vo:2010:SDD:1884643.1884681}, provide a coverage guarantee over the space of MPI non-determinism. When two or more matches of a non-deterministic operation, such as wildcard receive is detected, the program will be re-executed, and each running using a specific match. Hence, these tools can find the bug relying on a particular choice when non-deterministic operations are encountered, but also depend on the inputs that are fed to run the program. TASS \cite{TASS} tackles the limitation by using symbolic execution to reason about all the inputs of an MPI program, but its feasibility is limited by the simple MPI model used, which is justified in Section \ref{subSec:ee}. There are few static analysis work for MPI program. Greg proposes in \cite{DBLP:conf/cgo/Bronevetsky09} a novel data flow analysis notion, called \emph{parallel} control flow graph (pCFG), which can capture the interaction behavior of an MPI program with arbitrary number of processes. Based on pCFG, some static analysis activities can be carried out. 
However, the static analysis based on pCFG is hard to be automated.

Compared with the existing work, \mpise{} symbolically executes MPI programs and uses an on-the-fly scheduling algorithm to handle non-determinism, and provides the coverage guarantee on both input and non-determinism. In addition, \mpise{} uses a realistic MPI library, \emph{i.e.}, TOMPI \cite{Demaine97athreads-only}, to be the MPI model used. Therefore, more realistic MPI programs can be analyzed automatically by \mpise{}, without modifying the programs manually. Furthermore, since the MPI library and the symbolic executor are loosely coupled by hooking the library, it is not hard to switch to another implementation of MPI library to improve the precision and the feasibility of symbolic execution.
%
\section{Conclusion}
\label{sec:conclusion}

MPI plays a significant role in parallel programming. To improve the reliability of the softwares implemented using MPI, we propose \mpise{} in this paper to use symbolic execution to analyze MPI programs, targeting to find the bugs of an MPI program automatically. Existing work on analyzing MPI programs suffers problems in different aspects, such as scalability, feasibility and input or non-determinism coverage.
%
We employ symbolic execution to tackle the input coverage problem, and propose an on-the-fly algorithm to reduce the interleaving explorations for non-determinism coverage, while ensuring the soundness and completeness.
%
We have implemented a prototype of \mpise{} as an adoption of Cloud9, and conducted extensive experiments. The experimental results show that \mpise{} can find bugs effectively and efficiently. \mpise{} also provides diagnostic information and utilities to help people understand a bug.

For future work, there are several aspects. In one aspect, we plan to support non-blocking MPI operations, which are widely used in nowadays MPI programs. In another aspect, we want to refine our MPI model further, \emph{e.g.}, using a more realistic library, to improve the precision of symbolic execution. Finally, we are also concerned with improving the scalability of \mpise{} and the analysis of production-level MPI programs.
\bibliography{main}

\begin{thebibliography}{10}

\bibitem{DDT}
Allinea {DDT.}
\newblock \url{http://www.allinea.com/products/ddt/}.

\bibitem{GDB}
The {GNU} debugger.
\newblock \url{http://www.gnu.org/software/gdb/}.

\bibitem{NPB}
{NSA} parallel benchmarks.
\newblock \url{http://www.nas.nasa.gov/Resources/Software/npb.html}.

\bibitem{TotalView}
Totalview {Software.}
\newblock \url{http://www.roguewave.com/products/totalview}.

\bibitem{Baier:2008:PMC:1373322}
Christel Baier and Joost-Pieter Katoen.
\newblock {\em Principles of Model Checking (Representation and Mind Series)}.
\newblock The MIT Press, 2008.

\bibitem{DBLP:conf/cgo/Bronevetsky09}
Greg Bronevetsky.
\newblock Communication-sensitive static dataflow for parallel message passing
  applications.
\newblock In {\em Proceedings of the 7th Annual IEEE/ACM International
  Symposium on Code Generation and Optimization}, CGO '09, pages 1--12. IEEE
  Computer Society, 2009.

\bibitem{Bucur:2011:PSE:1966445.1966463}
Stefan Bucur, Vlad Ureche, Cristian Zamfir, and George Candea.
\newblock Parallel symbolic execution for automated real-world software
  testing.
\newblock In {\em Proceedings of the Sixth Conference on Computer Systems},
  EuroSys '11, pages 183--198, New York, NY, USA, 2011. ACM.

\bibitem{Cadar:2008:KUA:1855741.1855756}
Cristian Cadar, Daniel Dunbar, and Dawson Engler.
\newblock {KLEE}: Unassisted and automatic generation of high-coverage tests
  for complex systems programs.
\newblock In {\em Proceedings of the 8th USENIX Conference on Operating Systems
  Design and Implementation}, OSDI '08, pages 209--224, Berkeley, CA, USA,
  2008. USENIX Association.

\bibitem{Cadar:2011:SES:1985793.1985995}
Cristian Cadar, Patrice Godefroid, Sarfraz Khurshid, Corina~S.
  P\u{a}s\u{a}reanu, Koushik Sen, Nikolai Tillmann, and Willem Visser.
\newblock Symbolic execution for software testing in practice: Preliminary
  assessment.
\newblock In {\em Proceedings of the 33rd International Conference on Software
  Engineering}, ICSE '11, pages 1066--1071, New York, NY, USA, 2011. ACM.

\bibitem{Clarke:2000:MC:332656}
Edmund~M. Clarke, Jr., Orna Grumberg, and Doron~A. Peled.
\newblock {\em Model checking}.
\newblock MIT Press, Cambridge, MA, USA, 1999.

\bibitem{Demaine97athreads-only}
Erik Demaine.
\newblock A threads-only {MPI} implementation for the development of parallel
  programs.
\newblock In {\em Proceedings of the 11th International Symposium on High
  Performance Computing Systems}, pages 153--163, 1997.

\bibitem{xianghuadeng2006ase}
Xianghua Deng, Jooyong Lee, and Robby. Bogor/Kiasan.
\newblock A k-bounded symbolic execution for checking strong heap properties of
  open systems.
\newblock In {\em Proceedings of the 21st IEEE/ACM International Conference on
  Automated Software Engineering}, ASE '06, pages 157--166, 2006.

\bibitem{godefroid2008sage}
P.~Godefroid, M.Y. Levin, D.~Molnar, et~al.
\newblock Automated whitebox fuzz testing.
\newblock In {\em Proceedings of the Network and Distributed System Security
  Symposium}, NDSS '08, 2008.

\bibitem{Gopalakrishnan:2011:FAM:2043174.2043194}
Ganesh Gopalakrishnan, Robert~M. Kirby, Stephen Siegel, Rajeev Thakur, William
  Gropp, Ewing Lusk, Bronis~R. De~Supinski, Martin Schulz, and Greg
  Bronevetsky.
\newblock Formal analysis of {MPI}-based parallel programs.
\newblock {\em Commun. ACM}, 54(12):82--91, December 2011.

\bibitem{Hilbrich:2012:MRE:2388996.2389037}
Tobias Hilbrich, Joachim Protze, Martin Schulz, Bronis~R. de~Supinski, and
  Matthias~S. M\"{u}ller.
\newblock {MPI} runtime error detection with must: Advances in deadlock
  detection.
\newblock In {\em Proceedings of the International Conference on High
  Performance Computing, Networking, Storage and Analysis}, SC '12, pages
  30:1--30:11, Los Alamitos, CA, USA, 2012. IEEE Computer Society Press.

\bibitem{K76}
J.King.
\newblock Symbolic execution and program testing.
\newblock {\em Communications of the ACM}, 19(7):385--394, 1976.

\bibitem{KrammerBMR03}
Bettina Krammer, Katrin Bidmon, Matthias~S. M{\"u}ller, and Michael~M. Resch.
\newblock Marmot: An {MPI} analysis and checking tool.
\newblock In Gerhard~R. Joubert, Wolfgang~E. Nagel, Frans~J. Peters, and
  Wolfgang~V. Walter, editors, {\em PARCO}, volume~13 of {\em Advances in
  Parallel Computing}, pages 493--500. Elsevier, 2003.

\bibitem{CPE:CPE701}
Glenn~R. Luecke, Yan Zou, James Coyle, Jim Hoekstra, and Marina Kraeva.
\newblock Deadlock detection in {MPI} programs.
\newblock {\em Concurrency and Computation: Practice and Experience},
  14(11):911--932, 2002.

\bibitem{MPI2.2}
{Message Passing Interface Forum}.
\newblock {MPI}: A message-passing interface standard,version 2.2.
\newblock \url{http://www.mpi-forum.org/docs/}, 2009.

\bibitem{ita}
Patrick Ohly and Werner Krotz-Vogel.
\newblock Automated {MPI} correctness checking what if there was a magic
  option?
\newblock In {\em 8th LCI International Conference on High-Performance
  Clustered Computing}, South Lake Tahoe, California, USA, May 2007.

\bibitem{MPISPIN}
Stephen~F. Siegel.
\newblock Verifying parallel programs with {MPI}-spin.
\newblock In {\em Proceedings of the 14th European Conference on Recent
  Advances in Parallel Virtual Machine and Message Passing Interface}, PVM/MPI
  '07, pages 13--14, Berlin, Heidelberg, 2007. Springer-Verlag.

\bibitem{TASS}
StephenF. Siegel and TimothyK. Zirkel.
\newblock {TASS}: The toolkit for accurate scientific software.
\newblock {\em Mathematics in Computer Science}, 5(4):395--426, 2011.

\bibitem{Strout:2006:DAM:1156433.1157634}
Michelle~Mills Strout, Barbara Kreaseck, and Paul~D. Hovland.
\newblock Data-flow analysis for {MPI} programs.
\newblock In {\em Proceedings of the 2006 International Conference on Parallel
  Processing}, ICPP '06, pages 175--184, Washington, DC, USA, 2006. IEEE
  Computer Society.

\bibitem{cav08-isp}
Sarvani Vakkalanka, Ganesh Gopalakrishnan, and RobertM. Kirby.
\newblock Dynamic verification of mpi programs with reductions in presence of
  split operations and relaxed orderings.
\newblock In Aarti Gupta and Sharad Malik, editors, {\em Computer Aided
  Verification}, volume 5123 of {\em Lecture Notes in Computer Science}, pages
  66--79. Springer Berlin Heidelberg, 2008.

\bibitem{ppopp-VakkalankaSGK08}
Sarvani~S. Vakkalanka, Subodh Sharma, Ganesh Gopalakrishnan, and Robert~M.
  Kirby.
\newblock Isp: A tool for model checking mpi programs.
\newblock In {\em Proceedings of the 13th ACM SIGPLAN Symposium on Principles
  and Practice of Parallel Programming}, PPoPP '08, pages 285--286, New York,
  NY, USA, 2008. ACM.

\bibitem{conf/sc/VetterS00}
Jeffrey~S. Vetter and Bronis~R. de~Supinski.
\newblock Dynamic software testing of mpi applications with umpire.
\newblock In {\em Proceedings of the 2000 ACM/IEEE Conference on
  Supercomputing}, Supercomputing '00, Washington, DC, USA, 2000. IEEE Computer
  Society.

\bibitem{Vo:2010:SDD:1884643.1884681}
Anh Vo, Sriram Aananthakrishnan, Ganesh Gopalakrishnan, Bronis R.~de Supinski,
  Martin Schulz, and Greg Bronevetsky.
\newblock A scalable and distributed dynamic formal verifier for {MPI}
  programs.
\newblock In {\em Proceedings of the 2010 ACM/IEEE International Conference for
  High Performance Computing, Networking, Storage and Analysis}, SC '10, pages
  1--10, Washington, DC, USA, 2010. IEEE Computer Society.

\end{thebibliography}
\bibliographystyle{plain}
\newpage
\appendix
\section{Theorems and Proofs}
\label{append:models}

In order to model MPI programs, we introduce the notion of \emph{transition system} in \cite{Baier:2008:PMC:1373322}:
\begin{Definition}{Transition System}

A \emph{transition system} TS is a tuple $(S, Act,\rightarrow, I, AP, L)$, where
\begin{itemize}
\item $S$ is a set of states,
\item $Act$ is a set of actions,
\item $\rightarrow \subseteq S\times Act \times S$ is a transition relation,
\item $I\in S$ is a set of initial states,
\item $AP$ is a set of atomic propositions, and
\item $L:S\rightarrow 2^{AP}$ is a labeling function.
\end{itemize}
\end{Definition}

For an action $act$ and a state $s$, if there is a transition $\tau = \langle s, act, s'\rangle \in \rightarrow$, we say that $act$ is \emph{enabled} in state $s$, and $\tau$ is denoted as $s\overset{act}{\rightarrow} s'$. If there are no such $\tau$, $act$ is \emph{disabled} in $s$. We denote the set of actions enabled in $s$ as $enabled(s)$, \emph{i.e.}, $\{  act \mid  s_1\overset{act}{\rightarrow}s_2\wedge  s_1 = s\}$. Note that all transitions systems in this paper is assumed to be \emph{action deterministic}, \emph{i.e.} for a sate $s\in State$ and an action $\alpha \in Act$, $s$ has at most one transition labeled with $\alpha$ to another state. Hence if $\alpha\in Act$ is enabled in state $s$, we also use $\alpha(s)$ to denote the unique $\alpha$-successor of $s$, \emph{i.e.} $s\overset{\alpha}{\rightarrow}\alpha(s)$.

And we use \emph{execution} to describe the behavior of the transition system. An \emph{execution} in a transition system is an alternative sequence of states and actions $\pi = s_0\alpha_1 s_1\alpha_2\dots,\alpha_n s_n$ starts from a initial sate $s_0$ and ends at a terminal state, where $s_i\overset{\alpha_{i+1}}{\rightarrow} s_{i+1}$ holds for $ 0 \leq i < n$. We use $\lvert \pi \rvert$ to denote the \emph{length} of $\pi$, and $\lvert \pi \rvert = n$.

To model an MPI process, we define the model more precisely: $S = Loc\times Eval(Var)$ is a set of states, where $Loc$ is the locations of a process, and $Eval(Var)$ denote the set of variable evaluations that assign values to variables. $Act = \{s,r,b\}$, which means we only care about the blocking synchronous send and receive MPI operations as well as the collective operation barrier. We use $dest(op)$, where $op\in \{s,r\}$, to denote the destination of a send or a receive.

The above definition refers to only one process. However, the running of an MPI program typically consists of many processes. Therefore, we need mechanisms to provide the operational model for parallel runnings in terms of transition systems.

\begin{Definition}{Parallel composition}\\
Let $TS_{i}=(S_{i},Act_{i},\rightarrow_{i},I_{i},AP_{i},L_{i})\ i=1,2,\dots,n$ be $n$ transition systems. The transition system $TS = TS_1 \interleave_{H} TS_2 \interleave_{H} \dots \interleave_{H} TS_{n}$ is defined to be:
{$$TS = (S_1\times S_2\times \dots S_n, Act_g, \rightarrow, I_1\times I_2\times \dots I_n, AP_1\cup AP_2 \cup\dots AP_n, L_g)$$} where the transition relation $\rightarrow$ is defined by the following rule:
\begin{itemize}
\item for matched actions $\alpha,\beta \in H=\{s,r,b\}$ in distinct processes:
 $$\frac{s_i\overset{\alpha}{\rightarrow_i}s'_i \quad\wedge\quad s_j\overset{\beta}{\rightarrow_j}s'_j  \quad\wedge\quad match(\alpha, \beta)}{\langle s_1,\dots,s_i,\dots ,s_j,\dots s_n\rangle\xrightarrow{SR}{\langle s_1,\dots,s'_i,\dots ,s'_j,\dots s_n\rangle}}$$
here $match(\alpha,\beta)$ if and only if $(\alpha= s \wedge \beta=r) \vee (\alpha= r\wedge\beta= s)$, $dest(\alpha)=j$, and $dest(\beta)=i$, $SR$ is the compositional global action of s and r.

\item for matched actions $\alpha = b$ in distinct processes:
$$\frac{s_1\overset{\alpha}{\rightarrow_1}s'_1 \quad\wedge\quad \dots  s_i\overset{\alpha}{\rightarrow_i}s'_i \quad\wedge\quad \dots s_n\overset{\alpha}{\rightarrow_n}s'_n}{\langle s_1,\dots,s_i,\dots s_n\rangle\xrightarrow{B}{\langle s'_1,\dots,s'_i,\dots s'_n\rangle}}$$
Here B is the compositional global action of the local action $b$ of each process.
\end{itemize}
The labeling function is defined by $L_g(\langle s_1,\dots,s_i,\dots,s_n\rangle)= \bigcup_{1\le i\le n} L(s_i) $. And note that actions in $Act_g$ are also introduced in the above two rules.
\end{Definition}


The composition of transition systems gives a global view of a directed graph $G=(S, \mathcal{T} )$, where the nodes in $S$ are global states and an edge in $\mathcal{T}$ is a global transition with action $SR$ or $B$. Note that the idea behind our on-the-fly schedule is that, for a global state, we only choose some of the transitions to move on and discard the others. Hence, we only explore a subgraph. To describe this subgraph, we first introduce some notions here.

Given a global state $\sigma$ in a composed transition system, we fix a total order on actions enabled in $\sigma$ according to $weight(act)$, where $act\in enabled(\sigma) , \delta = \sigma\overset{act}{\rightarrow} \sigma'$, and  $$weight(act)=\left\{
                 \begin{array}{ll}
                   1, & \hbox{if $act = B$;} \\
                   i, & \hbox{if $act = SR$ , $\sigma=\langle s_1,\dots, s_n\rangle$, $\sigma'=\langle s_1,\dots, s'_i,\dots,s'_j,\dots s_n\rangle$.}
                 \end{array}
               \right.
$$

\emph{i.e.}, $act_1 < act_2$ \textrm{iff} $weight(act_1) < weight(act_2)$. When an action $act \in enabled(s)$ has the minimal weight, we say that $act$ ranks first in $enabled(s)$.

 \begin{Definition}
Let $\tilde{G} = (S, \tilde{\mathcal{T}})$ be a subgraph of $G$, where $\tilde{\mathcal{T}}$ is defined as follows:
\[\tilde{\mathcal{T}} = \bigcup_{s\in S}\{\tau \mid \tau=\langle s, act, s'\rangle \wedge act\ \text{ranks first in}\ enabled(s)\}.\]
 \end{Definition}

We can see that this $\tilde{G}$ is formed according to the on-the-fly schedule, in which we always schedule the active process with the smallest rank. Now, we can present the main theorem, which guarantees the completeness and soundness of the on-the-fly schedule.

\setcounter{theorem}{0}
\begin{theorem}
\label{theorem1}
Given an execution $\pi$ in $G$ from a global initial state $\sigma_0$ to a deadlocked global state $\sigma$, there exists an execution $T$ from $\sigma_0$ to $\sigma$ in $\tilde{G}$ such that $\lvert T\rvert = \lvert \pi \rvert$. And vice versa.
\end{theorem}
To prove the above theorem, we introduce some notions first.
An \emph{independent} relation $I\in Act \times Act$ is a relation, satisfying the following two conditions:\\
for any state $s \in S$,  with $\alpha,\beta \in enabled(s)$ and $\alpha \neq \beta$,
\begin{itemize}
\item \emph{Enabledness}     $\alpha\in enabled(\beta(s))$ , $\beta \in enabled(\alpha(s))$.
\item \emph{Commutativity}  $\alpha(\beta(s))=\beta(\alpha(s))$.

\end{itemize}
Recall that $\alpha(s)$ denotes the unique $\alpha$ successor of $s$, \emph{i.e.} if $ s\overset{\alpha}{\rightarrow} s'$ holds, then $s'=\alpha(s)$. The dependency relation $D$ is the complement of $I$, \emph{i.e.}, $D = Act \times Act - I$.

The method we constructing a subgraph $\tilde{G}$  from a graph $G$ actually falls into the \emph{ample set} method proposed in  \cite{Clarke:2000:MC:332656}, which expands only a part of the transitions at each state, for a path not considered by the method, there is an equivalent path with respect to the specification considered. Among the four conditions \textbf{C0}-\textbf{C3} for selecting $ample(s)\subseteq enabled(s)$, the first two are as follows:\\
$\mathbf{C0} \quad ample(s)=\emptyset$ if and only if $enabled(s)=\emptyset$.\\
$\mathbf{C1}$\quad Let $s_0\overset{\alpha_0}{\rightarrow}s_1,\dots,\overset{\alpha_{n}}{\rightarrow}s_n\overset{\alpha_{n+1}}
{\rightarrow}t$ be a finite execution fragment, if $\alpha_{n+1}$ depends on $\beta\in ample(s_0)$, then there exists an $\alpha_i\in ample(s_0)$, where $0\leq i < n+1$, \emph{i.e.}, along every execution in the full state graph $G$ that starts at $s$, an action appears in the execution fragment starts at $s$ which is dependent on an action in $ample(s)$ cannot be executed without an action in $ample(s)$ occurring first.

 \textbf{C2} makes sure that: when $ample(s)\neq enabled(s)$, the actions in $ample(s)$ do not change the label of states with respect to the verifying properties. Here the property we concerned with is deadlock-reachability. Since we commute independent actions by applying \emph{commutativity} of independent relation, we do not need \textbf{C2}.
\textbf{C3} ensures that a cycle in the reduced graph needs to satisfy some requirements. Owing to the acyclicity of the state space in out context, we do not need \textbf{C3}.

In our circumstance, we define $ample(s)$ as following:
$$ample(s) =\left\{
 \begin{array}{ll}
    \{B\} & \textrm{if $B\in enabled(s)$} ; \\
    \{SR\} & \textrm{else if $SR\in enabled(s)  \wedge SR$}\textrm{ ranks first in $enabled(s)$}.
 \end{array}
\right. $$
Note that for a state $s\in S$, if $enabled(s)\neq \emptyset$, $\lvert ample(s)\rvert = 1$, \emph{i.e.}, $ample(s)$ has only one element. Hence, $C_0$ surely holds.

To check whether \textbf{C1} holds on the full state graph generated by our schedule algorithm, we first introduce some properties of the full state graph. Clearly, according to the definition of parallel composition, only $SR$ actions can be enabled simultaneously at a global state, and the $SR$ actions enabled at a same state are independent.
So, given a execution fragment $\pi=s_0\overset{\alpha_0}{\rightarrow}s_1\dots,\overset{\alpha_{j-1}}{\rightarrow}s_j\overset{\alpha_j}{\rightarrow}s_{j+1}$ starting from $s_0\in S$, $\alpha_j$ depends on an action $\beta\in ample(s_0)$. We want to prove that there exists a $\alpha_k$, where $0 \leq k < j$, and $\alpha_k \in ample(s_0)$.  Recall the definition of dependent relation $D$, we know that $\alpha_j$ and $\beta$ should meet one of the following cases:
\begin{enumerate}
  \item $\nexists s\in S$ such that $\alpha_j,\beta\in enabled(s)$;
  \item for any state $s\in S$, if $\alpha_j,\beta\in enabled(s)$, then $\alpha_j\not\in enabled(\beta(s))$ or $\beta \not\in enabled(\alpha_j(s))$, \emph{i.e.}, either $\alpha_j$ disables $\beta$ or $\beta$ disables $\alpha_j$;
  \item for any state $s\in S$, if $\alpha_j,\beta\in enabled(s)$ and $\alpha_j\in enabled(\beta(s)) \wedge \beta \in enabled(\alpha_j(s))$, then $\alpha_j(\beta(s)) \neq \beta(\alpha_j(s))$, \emph{i.e.}, $\alpha_j$ and $\beta$ are not commutative.
\end{enumerate}

Because only $SR$ actions can be enabled simultaneously at a global state, both case 2 and case 3 cannot hold under our context. Therefore, only case 1 holds for the state graph generated by our method, \emph{i.e.}, $\alpha_j$ and $\beta$ should not be both enabled at any state. Based on this result, we can get \textbf{C1} holds in our context by contradiction. Assume that $\{\alpha_0\dots\alpha_{j-1}\}\cap ample(s_0) = \emptyset$ holds for the execution $\pi$. Because $\beta$ is enabled at $s_0$, $\alpha_0$ and $\beta$ are independent, hence $\beta$ is also enabled at $s_1$. In addition, $\alpha_1$ is also enabled at $s_1$ and $\alpha_1 \neq \beta$, so $\alpha_1$ and $\beta$ are also independent. In the same way, we can get that each $\alpha_i\ ( 0\leq i < j)$ and $\beta$ are independent. Thus, by using commutativity,  we can get $\beta$ and $\alpha_j$ are both enabled at $s_j$, which violates \mbox{case 1}. Hence, the condition \textbf{C1} holds.

\begin{proof}
One direction, because $\tilde{G}$ is a subgraph of $G$, the execution $T$ from $\delta_0$ to $\delta$ in $\tilde{G}$ is also an execution from $\delta_0$ to $\delta$ in $G$, hence we got an execution $\pi = T$, and $\lvert T \rvert = \lvert \pi \rvert$.

The other direction is a little more complex. The basic idea is to construct a corresponding execution in the subgraph gradually based on the $ample$ set of each state passed in $\pi$.

Let $\pi$ be an execution in $G$ from $\delta_0$ to $\delta$. We construct a finite sequence of executions $\pi_0,\pi_1,\dots,\pi_n$, where $\pi_0 = \pi$ and $n=\lvert \pi \rvert$. 
Each execution $\pi_i$ is constructed based on the before execution $\pi_{i-1}$. For example, $\pi_1$ is constructed from $\pi_0$, \emph{i.e.}, $\pi$, according to the first action execution in $\pi$. Thus, we want to prove that the last execution $\pi_n$ is an execution in the subgraph, and shares the same first and last states with $\pi$. We can prove it by presenting the construction method of each step. We decompose each $\pi_i$ into two execution fragments, \emph{i.e.}, $\pi_i = \eta_i\circ\theta_i$, where $\eta_i$ is of length $i$ and $\eta_{i}\circ\theta_{i}$ is the concatenation of the two execution fragments.

Assuming that we have constructed $\pi_0,\dots,\pi_i$, we now turn to construct $\pi_{i+1} = \eta_{i+1}\circ\theta_{i+1}$. Let $s_0$ be the last state of the execution fragment $\eta_i$ and $\alpha$ be the first action of $\theta_i$. Note that $s_0$ is also the first state of the execution fragment $\theta_i$, \emph{i.e.},
$$\theta_i = s_0 \overset{\alpha_0 = \alpha}{\longrightarrow}s_1\overset{\alpha_1}{\longrightarrow}s_2\overset{\alpha_2}{\longrightarrow}\dots s_{\lvert \theta_i \rvert}$$
There are two cases:\\
   \textbf{A.} $\alpha\in ample(s_0)$. Then $\eta_{i+1} = \eta_{i}\circ (s_0\overset{\alpha}{\rightarrow}{\alpha(s_0)})$ and $\theta_{i+1}=s_1\overset{\alpha_1}{\longrightarrow}s_2\overset{\alpha_2}{\longrightarrow}\dots s_{\lvert \theta_i \rvert}$.\\ 
    \textbf{B.} $\alpha\not\in ample(s_0)$.  Note that $s_{\lvert \theta_i \rvert} = \sigma$ is a deadlock state, hence no action can be enabled at $\sigma$. Therefore, for any action $\beta\in ample(s_0)$, some actions that appear in $\theta_i$ must be dependent on $\beta$. The reason is: if all the actions that appears in $\theta_i$ is independent of $\beta$, then all the actions in $\theta_i$ cannot disable $\beta$, hence $\beta$ would be enabled at $s_{\lvert \theta_i \rvert} = \sigma$, which violates the premiss that $\sigma$ is a deadlock state.

Therefore, for any action $\beta$ in $ample(s_0)$, we can find an action $\alpha_j$ that appears in $\theta_i$, and $\alpha_j$ depends on $\beta$. According to \textbf{C1}, there must exist an action $\beta' \in ample(s_0)$, such that $\beta'$ occurs before $\alpha_j$. Because there may exist multiple actions that are in $ample(s_0)$ and occur before $\alpha_j$, we take the first one, say $\alpha_k$ and $\alpha_k \in ample(s_0)$. So, $\alpha_k$ is the first one among the elements of $ample(s_0)$ that occur in $\theta_i$. Clearly, the actions before $\alpha_k$, \emph{i.e.}, $\alpha_0, ..., \alpha_{k-1}$, are independent with $\alpha_k$. Hence, we can construct the following execution by using the commutativity condition $k$ times: \\
 $\xi = s_0\overset{\alpha_k}{\rightarrow}\alpha_k(s_0)\overset{\alpha=\alpha_0}{\longrightarrow}\alpha_k(s_1)\overset{\alpha_1}{\longrightarrow}\dots \overset{\alpha_{k-1}}{\longrightarrow}\alpha_k(s_k)\overset{\alpha_{k+1}}{\longrightarrow}s_{k+2}\overset{\alpha_{k+2}}{\longrightarrow}\dots\overset{\alpha_{\lvert \theta_i \rvert -1 }}{\longrightarrow}s_{\lvert \theta_i \rvert}$.

In this case $\eta_{i+1}=\eta_i \circ (s_0\overset{\alpha_k}{\rightarrow}\alpha_k(s_0))$ and $\theta_{i+1}$ is the execution fragment that is obtained from $\xi$ by removing the first transition $s_0\overset{\alpha_k}{\rightarrow}\alpha_k(s_0)$.

Clearly, $\pi_i$ and $\pi_{i+1}$ share the same last state. So, $\pi_0 = \delta$ and $\pi_{n}$ share the same last state of the execution, namely $\pi_{n}$ is also an execution from $\delta_0$ to $\delta$ in $G$. In addition, according to the construction procedure, $\lvert \pi \rvert = \lvert \pi_{ n}\rvert$ holds. Most importantly, in execution $\pi_{n}$, for any $0\leq j< n$, such that $s_j\overset{\alpha_j}{\longrightarrow}s_{j+1}$, $\alpha_j\in ample(s_j)$ holds. Therefore, $\pi_{n}$ is also an execution from $\delta_0$ to $\delta$ in $\tilde{G}$, and we take this execution as $T$.\Qed
\end{proof}

To prove the correctness and soundness of our lazy matching algorithm, we need to deal with wildcard receives. Hence the rules of parallel composition of transition systems need to be refined. Instead of redefine $match$ to make it work with wildcard receives, we make a new rule for matched send and wildcard receives, to distinct it with source specific receives.
\begin{itemize}
\item for matched actions $\alpha,\beta \in H=\{s,r_{*}\}$ in distinct processes, where $r_*$ is the wildcard receive:
 $$\frac{s_i\overset{\alpha}{\rightarrow_i}s'_i \quad\wedge\quad s_j\overset{\beta}{\rightarrow_j}s'_j  \quad\wedge\quad match(\alpha, \beta)}{\langle s_1,\dots,s_i,\dots ,s_j,\dots s_n\rangle\xrightarrow{SR_*}{\langle s_1,\dots,s'_i,\dots ,s'_j,\dots s_n\rangle}}$$ here $match(\alpha,\beta)$ if and only if $\alpha= s \wedge \beta=r_* $,  $dest(\alpha)=j$, and $dest(\beta)=ANY$, $SR_*$ is the compositional global action of $s$ and $r_*$.
\end{itemize}
We also need to redefine the subgraph $\tilde{G}$ because we have a new kind of global transitions.

 \begin{Definition}
 Let $\tilde{\mathcal{T}}= \underset{s\in S}{\bigcup} subtran(s)$ , where $subtrans(s)$ is defined as:
 {\small
$$subtran(s) =\left\{
 \begin{array}{ll}
    \{s\overset{B}{\rightarrow}B(s)\} & \textrm{if $B \in enabled(s)$} ; \\
    \{s\overset{SR}{\rightarrow}SR(s)\} & \textrm{else if $SR \in enabled(s) \wedge$}
         SR\ \textrm{ranks first in $enabled(s)$}; \\
    \{s\overset{act}{\rightarrow}act(s)\}   & \textrm{else if $act\in enabled(s)$}.
  \end{array}
\right. $$}\\
  Let $\tilde{G_*} = (S,\tilde{\mathcal{T}})$, which is a subgraph of the full state graph.
 \end{Definition}

Clearly, we can see that $\tilde{G_*}$ is the subgraph we formed according to the on-the-fly schedule plus lazy matching. Accordingly, we define $ample(s)$ as:
$$ample(s) =\left\{
 \begin{array}{ll}
    \{B\} & \textrm{if $B\in enabled(s)$} ; \\
    \{SR\} & \textrm{else if $SR\in enabled(s)  \wedge SR$}\textrm{ ranks first in $enabled(s)$};\\
    enabled(s) & \textrm{other}.
 \end{array}
\right. $$

And the following theorem addresses the correctness and soundness of lazy matching algorithm:

\begin{theorem}
\label{theorem2}
Given any execution $\pi$ in $G$ from a global state $\sigma_0$ to a deadlocked global state $\sigma$, there exists an execution $T$ from $\sigma_0$ to $\sigma$ in $\tilde{G_*}$ such that $\lvert T\rvert = \lvert \pi \rvert$. And vice versa.
\end{theorem}

To prove theorem \ref{theorem2}, we first check the conditions \textbf{C0} and \textbf{C1}. Clearly, \textbf{C0} holds. To check \textbf{C1}, we should point out that in the new global state graph, an action $SR\in enabled(s)$ is independent with the rest actions in $enabled(s)$. In addition, only $SR_*$ can disable other $SR_*$ actions, which is ensured by the semantics of wildcard receives. 
Same as before, given a execution fragment $\pi=s_0\overset{\alpha_0}{\rightarrow}s_1\dots,\overset{\alpha_{j-1}}{\rightarrow}s_j\overset{\alpha_j}{\rightarrow}s_{j+1}$ starting from $s_0\in S$, $\alpha_j$ depends on an action $\beta \in ample(s_0)$. We want to prove that there exists a $\alpha_k$, where $0 \leq k < j$, and $\alpha_k \in ample(s_0)$.
We discuss the two cases of $ample(s_0)$:
\begin{enumerate}
  \item $ample(s_0) = enabled(s_0)$. Clearly, \textbf{C1} holds.
  \item $ample(s_0) \neq enabled(s_0)$. Thus, $ample(s_0) = \{SR\}$ and $\beta = SR$. 
We then discuss two cases: 1) $\alpha_j\in enabled(s_0) \setminus ample(s_0)$, according to the observation, $SR$ is independent with each of the rest actions in $enabled(s_0)$, so $\alpha_j$ and $SR$ are independent. Therefore, it is a contradiction, thus this case never happens. 2) $\alpha_j\not\in enabled(s_0)$. For an $SR$ action, the dependent relation of it can only be one case, \emph{i.e.}, $\nexists s\in S$ such that $\alpha_j,\beta\in enabled(s)$.
  Because $SR$ will never be disabled by any other action, same as the idea for proving \textbf{C1} for the case without wildcard receives, we can prove that $\beta$ occurs before $\alpha_j$.
\end{enumerate}
In total, we can get that \textbf{C1} holds on the new state graph.

\begin{proof}
We have concluded that the condition \textbf{C0} and \textbf{C1} still holds in the new full state graph. Hence, the procedure of the proof for theorem 2 is basically the same with that of theorem 1.
\Qed
\end{proof}

\end{document}